\documentclass[aps,pre,twocolumn,longbibliography,showpacs,amsmath,amsmath,amssymb,amsfonts,superscriptaddress]{revtex4-1}
\usepackage{graphicx}
\usepackage{dcolumn}
\usepackage{bm}
\usepackage{color}
\usepackage{multirow}
\usepackage{hyperref}
\usepackage{natbib}
\usepackage{xcolor}
\definecolor{myOrange}{rgb}{1,0.5,0}
\definecolor{myRed}{rgb}{0.8, 0.2, 0}
\definecolor{mygreen}{rgb}{0, 0.7, 0}

\usepackage{amsmath, amssymb}
\usepackage{bbold}

\begin{document}
\title{Generalized optimal paths and weight distributions revealed through the large deviations of random walks on networks}

\author{ Ricardo Guti\'errez}
\affiliation{Complex Systems Interdisciplinary Group (GISC), Department of Mathematics, Universidad Carlos III de Madrid, 28911 Legan{\'e}s, Madrid, Spain}
\author{Carlos P\'erez-Espigares}
\affiliation{Departamento de Electromagnetismo y F\'isica de la Materia, Universidad de Granada, Granada 18071, Spain}
\affiliation{Institute Carlos I for Theoretical and Computational Physics, Universidad de Granada, Granada 18071, Spain}

%\date{\today}
\begin{abstract}
Numerous problems of both theoretical and practical interest are related to finding shortest (or otherwise optimal) paths in networks, frequently in the presence of some obstacles or constraints. A somewhat related class of problems focuses on finding optimal distributions of weights which, for a given connection topology, maximize some kind of flow or minimize a given cost function. We show that both sets of problems can be approached through an analysis of the large-deviation functions of random walks. Specifically, a study of ensembles of trajectories allows us to find optimal paths, or design optimal weighted networks, by means of an auxiliary stochastic process (the generalized Doob transform).  In contrast to standard approaches, the paths are not limited to shortest paths and the weights must not necessarily optimize a given function. Paths and weights can in fact be tailored to a given statistics of a time-integrated observable, which may be an activity or current, or local functions marking the passing of the random walker through a given node or link. We illustrate this idea with an exploration of optimal paths in the presence of obstacles, and networks that optimize flows under constraints on local observables.
\end{abstract}

%\pacs{...}

\maketitle

\section{Introduction}

Shortest paths are one of the main objects of interest in network science, an interest driven by applications in transportation networks \cite{zhan1998,zeng2009,li2010}, the internet \cite{albert1999,tangmunarunkit2001,fortz2004,echenique2004,yan2006} and protein--protein interaction networks \cite{managbanag2008,jiang2013}, among other systems. They also enter into the definition of key structural  properties such as the closeness, the efficiency or the betweenness centrality  \cite{boccaletti2006, newman2010}. The definition of a shortest path between two nodes can be generalized to include, for example, constraints or obstacles (see e.g.\! \cite{zheng1996,carlyle2008,lozano2013,hershberger2020}) or to reach several targets from one or several sources (see e.g. \cite{knopp2007,jagadeesh2019}). A great deal of work in the applied discrete mathematics, theoretical computer science and statistical physics communities has dealt with the solutions of such problems, and the computational complexity of the algorithms involved to find them. 

Generally speaking, this research addresses problems such as how to find the shortest path between two nodes always passing, or without ever passing, through another one; but does not consider situations where a node is visited a fraction of times. Moreover, these methods cannot be easily extended to address problems where a given path observable is required to take a specific (not necessarily extremal) value, which may be characterized in a statistical sense. For example, it is not clear how one goes about finding the path or paths that maximize the fluctuations  (e.g. the variability across realizations of a stochastic process) of a given current; or the path that guarantees that a certain node is visited on average twice as frequently as another node.
%; or that a given target is reached from a source without going too frequently through a given obstacle. 
In more practical terms,  what is the shortest path that a passenger or a data packet can take without saturating a given node or link? Or what is the choice that maximizes the number of paths taken without the overall path length exceeding a certain threshold? These can also be considered optimal paths in a generalized sense, as they are chosen to ensure that the statistics of a given observable for a particle moving across the network takes specific values or does not exceed certain bounds. 

Similarly, one could think of redistributing the link weights of a network, characterized by a given adjacency matrix, so as to ensure that a set of nodes is visited with some frequency in the resulting weighted network, or that no link carries more than a certain amount of some flow. While this problem is less intensively investigated than that of optimal paths, examples can be found in the literature of e.g.\! networks that are optimal for sustaining a synchronized dynamics  \cite{tanaka2008,nishikawa2006,sevilla2016,kempton2017}. Recent contributions based on a kindred dynamical-control approach
show how to engineer force fields in many-particle systems so as to achieve prescribed steady-state distributions \cite{nemoto2017,ray18a,hurtadogutierrez2020}.

In this paper we propose a theoretical approach to unveil such generalized optimal paths and weight distributions by studying the statistics of trajectories using large-deviation methods \cite{lecomte2007,garrahan2009,touchette2009}. Specifically, we analyze the large deviations of random walks on graphs  \cite{debacco2016, coghi2019}. This allows us to find paths that are optimal in the statistical sense outlined above, or weight distributions that make a network optimal for a given statistical characterization pertaining to the flow of information or physical entities. By biasing the dynamics with certain observables we obtain the random-walk stationary distribution that guarantees that such observables (e.g.\! the activity associated with a node or link, or a current in a specific direction in a spatial network) satisfy some statistical constraint, which may be related to its mean value, fluctuations, or higher-order cumulants. We then employ the auxiliary process given by the generalized Doob transform \cite{simon09a,popkov10a,jack2010,chetrite2015}, to find the transition probabilities which give rise to the long-time statistics of the above-mentioned observable. By combining the biased stationary state with the generalized Doob transform, we extract the probability fluxes of the biased walk, which highlight the existence of optimal paths. Furthermore, the Doob-transformed process itself yields an optimal distribution of link weights in the same generalized sense.

We illustrate this versatile approach by finding generalized optimal paths in random graphs in the presence of constraints. To this end, we study the trajectories of the maximal entropy random walk \cite{burda2009} in an appropriate trajectory ensemble using as observable local activities. We also study constrained optimal weight distributions for maximal current or flow on spatial networks by means of the standard random walk \cite{noh2004}. These choices have been made as each is uniquely well suited to the problem under study. Once the choice of the appropriate process, observables and trajectory ensemble is made for a given problem involving generalized optimal paths or weight distributions, the solution can be found by a few steps of linear algebra. 

While the large-deviation approach lies at the basis of equilibrium statistical mechanics, and is  regarded as the natural language for dealing with many problems in non-equilibrium statistical physics \cite{touchette2009}, its application to the study of networks has started only recently. We conclude this introduction with a brief summary of some significant recent research. In the first work, as far as we are aware, on large deviations of time-integrated observables of random walks on networks \cite{debacco2016}, localization and mode-switching dynamical phase transitions are revealed.  More recently, in a contribution that has strongly influenced our methodology \cite{coghi2019},  such localization phenomena are explained by means of the generalized Doob transform, which is also used to shed new light on the relationship between the maximum entropy random walk and the standard random walk. Various other processes have been explored with related methodologies in publications dealing with, e.g., percolation transitions in single or multilayer networks subject to rare initial configurations, \cite{bianconi2018,bianconi2019}, paths leading to epidemic extinction \cite{hindes2016,hindes2017}, the connection between the rate of rare events and heterogeneity in population networks \cite{hindes2019}, or large-fluctuation-induced phase switch in majority-vote models \cite{chen2017}. Additionally, large-deviation and rare-event techniques have been employed in the exploration of  structural properties, such as the assortativity in configuration-model networks  \cite{chen2019}, the study of ensembles of random graphs satisfying structural constraints \cite{denhollander2018,giardina2020}, and the existence of a first-order condensation transition in the node degrees \cite{metz2019}.

\section{Thermodynamics of trajectories of random walks}

Random walks have been studied in continuous media and discrete spaces. Among the latter, much recent work has been devoted to the study of random walks on networks \cite{masuda2017,riascos2019}. Here we consider two different types of discrete-time random walks on networks: the standard random walk (SRW) \cite{noh2004} and the maximal entropy random walk (MERW) \cite{burda2009}. Both are discrete-time Markov chains whose components of the probability vector ${\bm p}$ evolve in time as $p_j(n+1) =  \sum_i \Pi_{ji}\, p_i(n)$, where the non-negative integer $n$ is the time step,  $p_i(n)$ is the probability that a random walker visits node $i$ at time $n$, and $\Pi_{ji}$ is the probability of a transition to $j$ conditioned on the node being in $i$. 
As usual, $\sum_i p_i(n) = 1$ for all $n$, and the probabilities of all possible transitions from a given node add up to one, $\sum_j  \Pi_{ji} = 1$. 

%We consider two types of random walks, namely, the standard random walk (SRW) and the maximal-entropy random walk (MERW). 
The SRW is suitable for the study of flow on networks ---currents, fluid flow, goods, etc.--- \cite{ahuja1993}, as it considers that a particle in a node can jump to any of its neighbors with the same probability (in the case of unweighted networks) or with probabilities proportional to the link weights (for weighted networks). Given an unweighted network ---this will be our starting point, though the generalization to weighted networks does not pose any difficulty--- with directed adjacency matrix ${\bf A}$, where $A_{ji} = 1$ if there is a link pointing from $i$ to $j$ and is $0$ otherwise, the entries of the transition matrix ${\bf \Pi}$ are 
\begin{equation}
\Pi_{ji}^{\text{SRW}} = \frac{A_{ji}}{k^\text{out}_i}.
\label{swr}
\end{equation}
The normalization by the out-degree, which is defined as the number of neighbors joined by outgoing links, $k^\text{out}_i = \sum_j A_{ji}$, ensures the conservation of probability.

The MERW, on the other hand, is most suitable for the exploration of generalized optimal paths in networks, as it assigns the same probability to all trajectories joining two given nodes that comprise the same number of steps (which the SRW does not do, see Appendix A). The transition matrix is given by 
\begin{equation}
\Pi_{ji}^\text{MERW} = \frac{A_{ji}}{\lambda}\frac{v_j}{v_i}.
\label{mewr}
\end{equation}
where $\lambda$ is the largest eigenvalue of the directed adjacency matrix, and ${\bm v}$ is the normalized eigenvector associated with it, ${\bm A}{\bm v} = \lambda {\bm v}$. For a more detailed discussion of the SRW and the MERW, as well as references treating other types of random walks on networks, see Appendix A.

For either type of random walk, a trajectory up to time $\tau$, $\omega_{\tau}$, is given by the sequence of nodes visited at each step, $\omega_{\tau}= (i_\tau \leftarrow \cdots \leftarrow i_2 \leftarrow i_1 \leftarrow i_0)$. We consider time-extensive observables of the form $\hat{O}(\omega_{\tau}) = \sum_{n=1}^\tau \hat{o}(i_n \leftarrow i_{n-1})$, with $ \hat{o}(i_n \leftarrow i_{n-1})$ being the increment of the observable at a time step, whose value depends on the nodes $i_{n-1}$ and $i_n$, which are joined by a link.  As the probability assigned to the trajectory is $P(\omega_{\tau}) =\Pi_{i_\tau i_{\tau-1}} \cdots \Pi_{i_2 i_1}\Pi_{i_1 i_0}p_{i_0}(0)$, %where $\Pi_{ji}$ is the transition matrix of the random walk under study, 
the probability distribution of the observable is $P_{\tau}(O) = \sum_{\omega_{\tau}}\delta(O - \hat{O}(\omega_{\tau})) P(\omega_{\tau})$. This distribution corresponds to an ensemble of trajectories with fixed observable $O$ and fixed time $\tau$. In the long-time limit, it adopts a large-deviation form $P_{\tau}(O)\sim e^{-\tau I(O/\tau)}$, which is here given in terms of the time-intensive observable $O/\tau$ (see Appendix B for details). The function $I(O/\tau)$ is called the rate function, which plays the role of a dynamical entropy. This ensemble of trajectories is analogous to the micro-canonical ensemble of equilibrium statistical mechanics, and it is generally speaking not the most useful one to work with. Fortunately, the thermodynamic formalism of time-integrated dynamical observables developed in Refs.~\cite{lecomte2007,garrahan2009} shows how to study the statistics of $O$ in more suitable ensembles.

By biasing trajectories with a parameter $s$ ---which we refer to as the \emph{tilting parameter}--- we obtain the $s$-ensemble, where $\tau$ is fixed but $O$ is now a fluctuating observable whose probability distribution is given by $P_{\tau}^s(O) = Z_{\tau}^{-1}(s)\,  e^{-s O}P_{\tau}(O)$. The normalization factor is a dynamical partition function which also acquires a large-deviation form for large $\tau$, $Z_{\tau}(s) = \sum_O e^{-s O} P_{\tau}(O) \sim e^{\tau \theta(s)}$, where $\theta(s)$ is the so-called scaled-cumulant generating function (SCGF), and is related to the rate function by a Legendre-Fenchel transform \cite{touchette2009}.  Its role is that of a dynamical free energy, and in fact from its derivatives one can obtain the cumulants of the time-intensive observable $O/\tau$ (see Appendix B). Crucially, the partition function can be expressed as $Z_{\tau}(s) = \sum_{i,j}  [\left({\bm \Pi^s}\right)^\tau]_{ji} p_{i}(0)$, where ${\bm \Pi^s}$ is the \emph{tilted operator}, which is akin to a transfer matrix, whose elements are
\begin{equation}
 \Pi_{ji}^s = e^{-s \hat{o}(j \leftarrow i)} \Pi_{ji}\,.
\label{tilteds}
\end{equation}
Thus, finding the SCGF $\theta(s)$ ---which fully characterizes the statistics of $O$--- reduces, in the long-time limit, to an eigenvalue problem for the tilted operator \eqref{tilteds}. Actually, by spectral decomposition it is straightforward to check that, for long times, ${\theta(s)}$ is given by the logarithm of the largest eigenvalue of ${\bm \Pi^s}$. %and $e^{\theta(s)}$ is its largest eigenvalue \cite{garrahan2009}. 
While the derivatives of $\theta(s)$ at $s=0$ correspond to the cumulants of $O$ in the typical (unbiased) distribution, $P_{\tau}(O)$, those evaluated at $s\neq 0$ provide information about the statistics of $O$ in the tilted (biased) one, $P^s_{\tau}(O)$. For $s\neq 0$ such statistics is carried by the so-called rare trajectories ---as they are exponentially unlikely---, which are not, however, easily retrieved from the tilted operator \eqref{tilteds}, as it is not a stochastic matrix ($\sum_j \Pi_{ji}^s \neq 1$). Nevertheless, as we later explain, this unphysical tilted operator can be transformed into a proper stochastic matrix by means of the generalized Doob transform, revealing the rare trajectories of interest. The latter arise from the transition probabilities leading to the fluctuation conjugate to the tilting parameter $s$.

%Finding the statistics of $O$, which here will be always a local (signed) activity that equals one (or minus one) if a given link (or set of links) $j \leftarrow i$ is traversed and zero otherwise, thus becomes an eigenvalue involving an operator that is not stochastic, $\sum_j \Pi_{ji}^s \neq 0$ for $s\neq 0$. While for $s\neq 0$ the dynamics is thus unphysical, the derivatives of $\theta(s)$ for $s\neq 0$ reveal the statistics of tilted generators that can be transformed into proper stochastic generators, and reveal the transition probabilities that give rise to them, as we explain below.

\begin{figure*}[t!]
\includegraphics[scale=0.147]{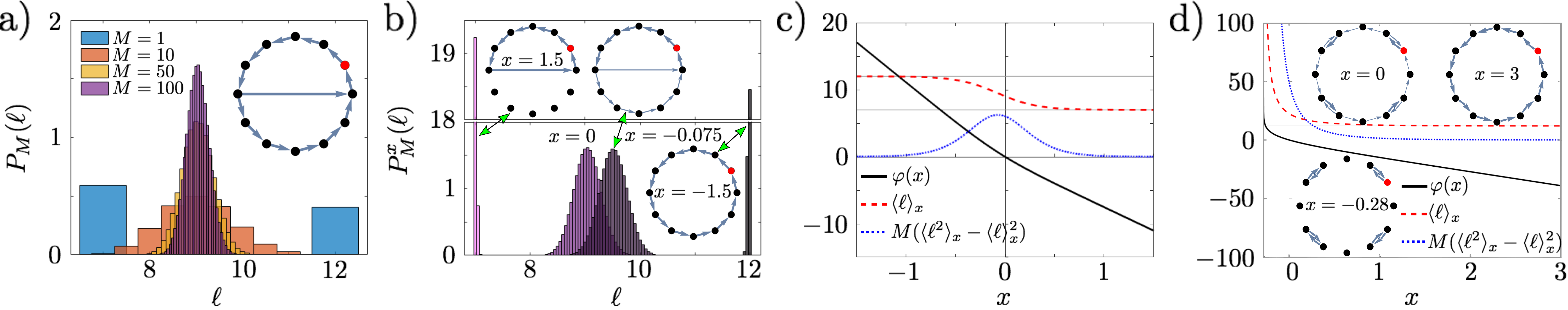}\\
\caption{{\sf \bf Finding paths that extremize the cycle length or maximize its fluctuations in simple graphs.}
(a) Probability distribution of the sample mean of the cycle length $\ell$ (see text for the definition) of $M$ random walk trajectories on the graph depicted on the upper right corner, starting and ending at the node highlighted in red, for $M=1,10,50,100$. The distribution follows a large deviation principle $P_M(\ell)\sim e^{-M I(\ell)}$ for large $M$, where only small fluctuations around the mean are Gaussian, since $I(\ell)$ is not quadratic for large fluctuations. (b) Original $\ell$ distribution ($x=0$) for $M=100$, and three tilted distributions: while $x = -0.075$ maximizes the fluctuations, $x=-1.5$ and $x=1.5$ peak around the longest and shortest possible lengths, respectively. The graphs display the probability fluxes as given by the Doob transform (arrow thickness is proportional to the probability flux). Probability fluxes smaller than the largest value divided by 200 are not displayed for visibility reasons.  (c) SCGF $\varphi(x)$ and path-length mean $\langle \ell \rangle_{x} = -\varphi^\prime(x)$ and (scaled) fluctuations $M (\langle \ell^2 \rangle_{x} - \langle \ell \rangle_{x} ^2) = \varphi^{\prime\prime}(x)$ for large $M$. (d) SCGF $\varphi(x)$, $\langle \ell \rangle_{x} $ and $M (\langle \ell^2 \rangle_{x}  - \langle \ell \rangle_{x}^2)$, and probability fluxes for $x=3$ (shortest path) and $x=-0.28$ (localized walker), corresponding to the graph depicted in this panel for $x = 0$.
}\label{fig1}
\end{figure*}

If instead of keeping fixed the duration of a trajectory $\tau$, we fix the value that the observable $O$ reaches for each trajectory ---thus allowing $\tau$ to fluctuate---, we obtain a different statistical ensemble, namely, the $x$-ensemble \cite{budini2014}. In this ensemble, we shall consider as observable $O$ a local activity, that is, the number of times a given link (or set of links) is traversed in a trajectory, having ${\hat o}(j \leftarrow i)=1$ whenever this occurs and zero otherwise. We denote by $P(y_O)$ the probability of a trajectory $y_O=(i_\tau \leftarrow \cdots \leftarrow i_2 \leftarrow i_1 \leftarrow i_0)$ that reaches a fixed value $O$ in a number of steps $\tau$. As the latter fluctuates from trajectory to trajectory, the probability distribution of the time duration for fixed $O$ is $P_O(\tau)= \sum_{y_O}\delta(\tau - \hat{\tau}(y_O)) P(y_O)$, where the operator $\hat{\tau}$ counts the number of time steps in a trajectory.
%The probability that the observable reaches a given value $O$ at time $\tau$ is the probability $P_O(\tau)$ that the selected links are traversed at times $\tau_1, \tau_2,\ldots,\tau_{O-1}, \tau_O = \tau$, where all but the last one can take any value. 
In this case, the dynamical partition function conditioned on a fixed value of $O$ reads
\begin{equation}
Z_O(x) =\sum_\tau e^{-x \tau} P_O(\tau) = \sum_{i,j} \left[({\bm \Pi}^x)^O\right]_{ji}p_{i}(0),
\end{equation}
which again we write in terms of a tilted operator, namely,
\begin{equation}
{\bm \Pi^x} = {\bf \Pi}_O (e^{x}- {\bf \tilde \Pi} )^{-1}.
\label{tiltedx}
\end{equation}
Here, ${\bf \Pi}_O$ is a matrix which preserves the transition probabilities of ${\bf \Pi}$ only for those transitions (links) contributing to $O$ (i.e.\! $j \leftarrow i $ such that ${\hat o}(j \leftarrow i) \neq 0$), the rest of its entries being zero, and  ${\bf \tilde \Pi}={\bf \Pi}-{\bf \Pi}_O$. For large $O$ (which also corresponds to large $\tau$), the grand-partition function acquires a large-deviation form $Z_O(x) \sim  e^{O \varphi(x)}$, where $e^{\varphi(x)}$ is the largest eigenvalue of ${\bm \Pi^x}$. The cumulants of the fluctuating time between observable $O$ updates, $\tau/O$, can then be obtained from the derivatives of the $x$-ensemble SCGF $\varphi(x)$ (see Appendix B).

While the $s$-ensemble and the $x$-ensemble are equivalent in the $\tau\to\infty$ (hence $O\to\infty$, as $O$ is time-extensive) limit  \cite{budini2014,garrahan2017}, the former is more natural for the study of time-averaged observables of the form $O/\tau$, and the latter is more appropriate for the analysis of their reciprocal, $\tau/O$. In the following, we will use one or the other depending on the specific problem under study. 

In fact, we will also consider ensembles of biased trajectories with two different tilting parameters. The latter will be conjugate to two fluctuating time-extensive observables, or a fluctuating observable and the duration of the trajectory $\tau$. A description of such $ss$- and $sx$-ensembles, as they are respectively referred to, as well as a more detailed characterization of the ensembles presented above can be found in Appendix B. In the $ss$- and $sx$-ensembles it will also be possible to obtain the relevant SCGFs by computing the largest eigenvalues of certain transfer operators, which are extensions of those given in Eqs.~(\ref{tilteds}) and (\ref{tiltedx}).

All such tilted operators, as explained above for the $s$-ensemble and regardless of the ensemble under consideration, have something in common: they are not stochastic operators. However, by an application of the generalized Doob transform \cite{simon09a,popkov10a,jack2010,chetrite2015} ---which for convenience we will just refer to as the Doob transform in the remainder of this paper---, one  obtains an auxiliary process whose statistics is given by $P^s_{\tau}(O)$ in the long-time limit. The Doob transform of a tilted operator is discussed in Appendix C and references therein. The transformed transition matrix ${\bm \Pi}_\text{Doob}$, which unlike the tilted operator is a proper stochastic matrix, gives us the set of transition probabilities that characterize the auxiliary process. By multiplying these with the corresponding stationary probabilities ${\bm p}^{\text st}$, which satisfy ${\bm \Pi}_\text{Doob}{\bm p}^{\text st}={\bm p}^{\text st}$, 
we obtain the probability fluxes $(\Pi_\text{Doob})_{ji}p^{\text st}_i$, which are the joint probabilities of being at node $i$ at a given time step, and moving to node $j$ at the next one. While the transition matrix ${\bm \Pi}_\text{Doob}$ will give us the optimal link weights in order to sustain a given statistics, the probability fluxes will visually reveal most clearly the network that results from imposing such statistics on the observables of interest, and highlight the optimal paths in it.

\section{Optimal paths from large deviations of maximal-entropy random walks}

\subsection{{\sf \bf Biased trajectories in directed rings }}

We first illustrate the idea of searching for optimal paths with a simple example. Consider the ring with a shortcut shown in Fig.~\ref{fig1} (a). A particle starts from the node which has been highlighted in red, and then hops counterclockwise to the neighboring node, and then hops again %and from there to the next one, and so on, 
always following the MERW transition probabilities. These are all identically one, except when the particle reaches the node from which the shortcut starts, where it may continue along the ring with probability $p \approx 0.41$ or it may take the shortcut with probability $q = 1-p \approx 0.59$. 

We will be interested in the statistics of the length (number of time steps) of a cyclic walk that starts and ends in the red node. If the cycle is performed $M$ times, we consider the probability distribution of the sample mean $\ell = M^{-1} \sum_{i=1}^M \ell^{(i)}$, where $\ell^{(i)}$ is the cycle length of a given realization. For a single cycle ($M=1$), the walker can either follow along the ring with probability $p$, which gives $\ell = \ell_p = 12$, or take the shortcut with probability $q$, for which $\ell = \ell_q = 7$, see  Fig.~\ref{fig1} (a). As $M$ grows, $\ell$ takes more and more values, and for sufficiently large $M$, the distribution centers around the mean $p\, \ell_p + q\, \ell_q \approx 9.04$ following a large deviation principle $P_M(\ell)\sim e^{-M I(\ell)}$, with fluctuations that are approximately Gaussian only around the mean --- i.e. $|\ell-\langle \ell \rangle|\sim{\cal O}(1/\sqrt{M})$---, as expected from the law of large numbers and the central limit theorem. Fluctuations that deviate far from the average are, however, not Gaussian and they are the prime concern of large-deviation theory.

In order to unveil optimal paths, we shall focus on the probability of large deviations of $\ell$. These are studied in the $x$-ensemble, since we are interested in the fluctuations of the length (number of time steps) for a given number of cycles.  The duration of the trajectory $\ell M$ is thus fluctuating (this would correspond to $\tau$ in Section II) and the number of realizations $M$ is fixed (this would correspond to $O$ in Section II). The latter condition can be achieved by fixing to $M$ the activity through the link that reaches the red node from the preceding node in the ring ---this is the local observable. We thus take the probability distribution $P_M(\ell)$, corresponding to $M=100$ cycles, and bias it using $x$ as tilting parameter,
\begin{equation}
P_M^x(\ell) =  e^{-x M \ell}  P_M(\ell)/Z_M(x).
\end{equation}
Here $P_M(\ell) = P_M^{x=0}(\ell)$, $Z_M(x)$ is a normalizing factor, and the large-deviation regime corresponds to large values of $M$. In  Fig.~\ref{fig1} (b), we show such tilted probability distribution for $M=100$ and different values of $x$, %$P_x(\ell_{100}) = e^{-x 100\, \ell} P(\ell_{100})/Z(x)$, where $Z(x) = \sum_\ell e^{-x 100 \ell} P(\ell_{100})$ 
namely, $x = -1.5,-0.075$ and $1.5$ ---the untilted case $x = 0$, which was already shown in panel (a), is also included for comparison. These choices of $x$ correspond to the largest ($x = -1.5$) and smallest ($x = 1.5$) possible $\langle \ell \rangle_x$, which are of course $\ell_p$ and $\ell_q$ respectively, and to the value ($x = -0.075$) that maximizes the fluctuations $\langle \ell^2 \rangle_x -\langle \ell \rangle_x^2$. This information is obtained from the SCGF $\varphi(x)$ of the $x$-ensemble of MERW trajectories, which is shown, together with its first and second derivatives,  in Fig.~\ref{fig1} (c). Such derivatives (with the appropriate signs) correspond to the mean and the (scaled) fluctuations of $\ell$: $\langle \ell \rangle_x = -\varphi^\prime(x)$ and $M (\langle \ell^2 \rangle_x - \langle \ell \rangle_x^2) = \varphi^{\prime\prime}(x)$.

In this simple directed ring, the analytical expression for the SCGF $\varphi(x)$ can be readily obtained from the large deviations of $\ell$, so one does not need to compute the largest eigenvalue of the tilted operator in Eq.~(\ref{tiltedx}). We briefly review the main steps of the calculation ---a detailed derivation can be found in Appendix D. As each $\ell^{(i)}$ is a Bernoulli trial which takes the values $\ell_p$ and $\ell_q$ with probabilities $p$ and $q=1-p$, the random variable $n_p$, quantifying the fraction of times that the path of length $\ell_p$ is taken, has a binomial probability distribution. Its large-deviation form can be obtained through an application of Stirling's approximation. By a change of variable we find the distribution of $\ell = n_p \ell_p + (1-n_p) \ell_q$, which also acquires a large-deviation form $P_M(\ell) \sim e^{-M I(\ell)}$. From the rate function $I(\ell)$, the SCGF $\varphi(x) = \lim_{M\to\infty} \log[Z_M(x)]/M$ is obtained via a Legendre transform $\varphi(x) = -\text{min}_\ell[x \ell+ I(\ell)]$, yielding
\begin{equation}
\varphi(x) = -x \ell_q + \log[p e^{-x(\ell_p -\ell_q)} + (1-p) ].
\label{varphidirring}
\end{equation} The analytical expression of its first derivative shows that the average $\langle \ell \rangle_x$  is bounded between $\ell_q$ and $\ell_p$, and approaches those bounds for large tilting-parameter (absolute) values.  The asymptotic value $\ell_q$ is reached for positive $x$, while $\ell_p$ corresponds to negative $x$, as expected for a sufficiently strong bias towards shorter/longer cycles. In Fig.~\ref{fig1} (c), for the values $p$, $\ell_p$ and $\ell_q$ under consideration, such extreme values are already practically reached  for $|x| \approx 1$, but this value will change if the shortcut is located elsewhere (see Appendix D).

The probability fluxes corresponding to those same values for which we show the tilted distributions, namely,  $x = -1.5,-0.075$ and $1.5$, are also displayed in  Fig.~\ref{fig1} (b).  They highlight the optimal paths in each of the three situations considered. The extreme values of $x$ correspond to a walk that just moves along the ring (for negative $x$, which favors long paths, $\langle \ell \rangle_x \approx \ell_p$) or takes the shortcut (for positive $x$, which favors short paths, $\langle \ell \rangle_x \approx \ell_q$). On the other hand, the maximization of the fluctuations leads to the shortcut being taken or avoided with probability $1/2$. While one can calculate the statistics of $\ell$ for all $x$ without resorting to the $x$-ensemble tilted generator (\ref{tiltedx})---at least for very simple systems whose SCGF $\varphi(x)$ can be found analytically---, the eigenvectors of such operator are still needed to compute the Doob transform, and therefore to obtain the probability fluxes, see Appendix C. Moreover, the distributions for  $|x|=1.5$ displayed in Fig.~\ref{fig1} (b) are also based on the Doob-transformed process, as this allows us to circumvent the challenging task of numerically performing very strong tiltings, which result in distributions lying on regions where $P_M(\ell)$ is negligibly small ---see the discussion of an analogous issue in the last section of Ref.~\cite{Carollo2018}.

While in the illustrative example we have just considered the resulting optimal paths are trivial, in more complex topologies our approach unveils paths and weight distributions whose adequacy for sustaining given observable statistics is far from obvious. But before moving on to such more interesting examples, let us illustrate the point with another simple case, namely, a ring with alternate bidirectional links, see Fig.~\ref{fig1} (d), where the original probability fluxes are shown for $x=0$. This allows us to discuss localization phenomena and diverging times, which will also appear later. In this case we compute the SCGF from the corresponding $x$-ensemble operator, and obtain from its derivatives the mean cycle length $\langle \ell \rangle_x$ and the (scaled) fluctuations $M(\langle \ell^2 \rangle_x - \langle \ell \rangle_x^2)$. For $x=3$ the probability fluxes show the shortest possible path, as expected, which moves counterclockwise along the ring with vanishingly small fluctuations. For negative $x$, however, there is a vertical asymptote in the SCGF, hence the cumulants diverge. The probability fluxes for a value of $x$ sufficiently close to the divergence show a situation where the particle becomes localized and never reaches the red node from its clockwise neighbor ---see the trajectories for $x=-0.28$ in Fig.~\ref{fig1} (d). Quite appropriately, $\langle \ell \rangle_x$ grows unboundedly as $x$ approaches the divergence and the particle becomes more and more localized. The mathematical origin of divergences in the $x$-ensemble is discussed in Appendix B.

\subsection{Constrained optimal paths in random graphs} 
\begin{figure}[t!]
\includegraphics[scale=0.150]{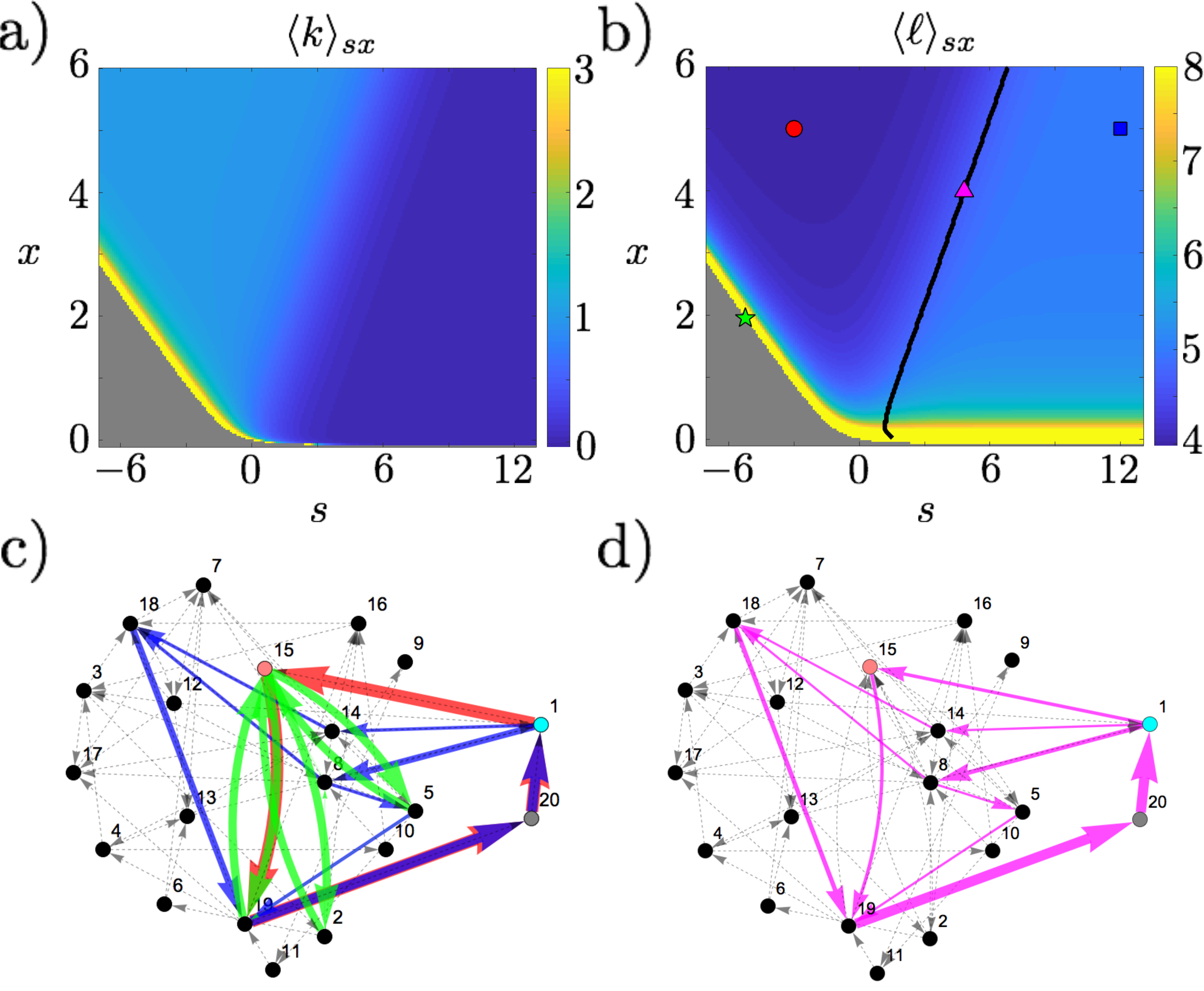}\\
\caption{{\sf \bf Finding optimal paths in random graphs in the presence of constraints.}
(a) Average length $\langle l \rangle_{sx}$ of path joining nodes $1$ (source) and $20$ (target) in the graph shown in panels (c) and (d) as a function of the tilting parameters $s$ and $x$. (b) Average activity $\langle k \rangle_{sx}$, i.e.\! number of times that node $15$ is visited before the target is reached, as a function of the tilting parameters $s$ and $x$. Highlighted points correspond to probability fluxes shown in the panels below. The black segment shows the contour line for $\langle k \rangle_{sx} = 1/3.$ (c) Probability fluxes obtained from the Doob-transformed process corresponding to the red circle, the green star and the blue square points shown in (b). (d) Probability fluxes obtained from the Doob transformed process of the magenta triangle in (b). Probability fluxes smaller than the largest value divided by 200 are not displayed for visibility reasons.
}\label{fig2}
\end{figure}

A more interesting case is considered next, namely, that of finding optimal paths in the presence of constraints in random graphs. To provide a concrete illustration, we consider a random graph of $N=20$ nodes with $3 N$ directed links distributed uniformly at random among them, see  Fig.~\ref{fig2} (c) and (d) ---the links highlighted in colors others than gray will be discussed below. Much larger networks with Poissonian or power-law degree distributions, or networks arising from applications that do not correspond to a precise mathematical model could be similarly studied.

In this case, a particle repeatedly performs a MERW from the source node $1$ and reaches the target node $20$ after $\ell$ steps. Again, we study the statistics of $\ell$, but this time we also consider whether the particle goes through an obstacle, node $15$, before reaching the target node. Here, ``obstacle'' is used in a loose sense to indicate that some constraint, based on how frequently the walker goes through that node, will be applied.  To this end, we employ a second observable, namely $k$, which gives the activity of the obstacle, defined as the number of times the particle goes through the obstacle before reaching the target node (see Ref.\cite{angeletti16a} for a similar observable in a diffusion process). The obstacle can be completely avoided, or it can just be avoided a fraction of the times that the target node is reached, or the particle may even be biased to visit it more frequently than in the natural dynamics ---it all depends on the value of the tilting parameter $s$ conjugate to $k$ (see below).  

Before discussing the results of our analysis, we mention a technical point which may be relevant in applications. The large-deviation analysis allows us to access the average, fluctuations and higher cumulants of $k$ and $\ell$, including correlations between them, for different tilting-parameter values. This is made possible by the existence of a link connecting node $20$ with node $1$, which restarts the process when the target node is reached and guarantees the time-extensivity of integrated observables  (see Appendix B for details). In practice, this may already be part of the network or, if not, it should be expressly introduced for the analysis. 

We consider an $sx$-ensemble, with tilted probability distribution
\begin{equation}
P^{s x}_M(k,\ell) =  e^{-s M k -x M \ell}  P_M(k,\ell)/Z_M(s,x)
\end{equation}
where $P_M(k,\ell) = P^{s=0,x=0}_M(k,\ell)$, $Z_M(s,x)$ is a normalizing factor, and $M$ is assumed to be fixed and large. The fluctuating time is again $M \ell$, and $M k$ is an $M$-extensive ---hence time-extensive--- observable (it corresponds to $K$ in Appendix B). The SCGF $\varphi(s,x)$ is calculated from the largest eigenvalue of the tilted operator (see Appendix B), and from its partial derivatives we obtain the average activity $\langle k \rangle_{sx} = -\partial_s \varphi(s,x)$ and the average path length $\langle \ell \rangle_{sx} = -\partial_x \varphi(s,x)$, which are shown in  Fig.~\ref{fig2} (a) and (b) respectively. The gray area on the lower left-hand corner corresponds to a region where the averages grow extremely rapidly and eventually diverge for reasons analogous to those given above regarding the divergence shown in Fig.~\ref{fig1} (d). For $x=0$ (no tilting is applied on $\ell$) the physical meaning is clear: as $s$ becomes negative, the particle goes through the obstacle more and more frequently before reaching the target node, and when $s$ is sufficiently large and negative, it never reaches it. For sufficiently large and positive $s$, on the other hand, the particle completely avoids the obstacle, so we have $\langle k\rangle_{sx} \approx 0$ and a finite $\ell$. As $x$ is increased (when the tilting favors shorter walks joining the source node and the target node), the particle goes more rapidly towards the target, and therefore it requires a larger negative value of $s$ to start growing steeply and eventually diverge. 

The probability fluxes derived from the Doob transform for the point ($s=-5.24$, $x =1.95$), which lies very close to the divergence and is highlighted with a green star in Fig.~\ref{fig2} (b), are shown as green arrows in panel (c). They clearly display a trajectory where the particle moves back and forth between the obstacle and its neighbors, without ever reaching the target. This and other points discussed below are only highlighted in panel (b), where the color map shows values of $\langle \ell \rangle_{sx}$, but of course they correspond to the same parameter-space points in (a), where the color map shows values of $\langle k\rangle_{sx}$. 

More importantly, a crossover between a region where  $\langle k\rangle_{sx} \approx 0$ and a plateau where  $\langle k\rangle_{sx} \approx 1$ is observed in (a), which corresponds to a crossover from a region where $\langle \ell \rangle_{sx}$ is larger than $4$ to a region where $\langle \ell \rangle_{sx} \approx 4$ in (b). The probability fluxes for the point ($s=-3$, $x =5$), which is highlighted with a red disk in Fig.~\ref{fig2} (b) and corresponds to $\langle \ell \rangle_{sx} = 4$, are shown as red arrows in panel (c). They show a trajectory where the particle moves along the shortest path between the source and the target, and gets back to the source after precisely four steps. As the obstacle lies on this path, we have  $\langle k\rangle_{sx} = 1$. On the other side of the crossover and for sufficiently large $s$ and $x$, for example for  ($s=12$, $x =5$), the walker chooses the shortest amongst the paths that avoid the obstacle, see the blue square in (b) and the blue arrows in (c).

One may be interested in finding the path or combination of paths that only visit the obstacle with a given frequency while reaching the target node in the smallest possible number of steps. 
In contrast with the cases discussed above, this type of generalized path is, as far as we are aware, outside the reach the graph-theoretical methods typically used for finding shortest paths, despite its clear practical interest, e.g.\! in transportation networks where a given station or airport may have a limited capacity that cannot be exceeded. To this end we set the activity of the obstacle to $\langle k \rangle_{sx} = 1/3$: one in three times when the target is reached the walker has passed through the obstacle. We then obtain the set of points of the $(s,x)$ grid where $\langle k \rangle_{sx} = 1/3$ with an error of $\pm 0.001$. The resulting segment is highlighted in black in  Fig.~\ref{fig2} (b). For $x \geq 4$ the average path length evaluated along the segment practically reaches an asymptotic value of $\langle \ell \rangle_{sx} \approx 4.67$, which is then the shortest time it takes to reach the target while passing through the obstacle only 1/3 of the times.  The probability fluxes for the point ($s=4.815$, $x =4$), which is highlighted with a pink triangle in Fig.~\ref{fig2} (b), are shown in panel (d). Apart from the shortest path, which occurs with a probability $1/3$ ---as it should, given that it contains the obstacle---, the other 2/3 are equally split among the three second shortest paths, out of a total of five, that do not cross the obstacle.

\section{Optimal weight distribution from large deviations of standard random walks}

\begin{figure}[t!]
\includegraphics[scale=0.133]{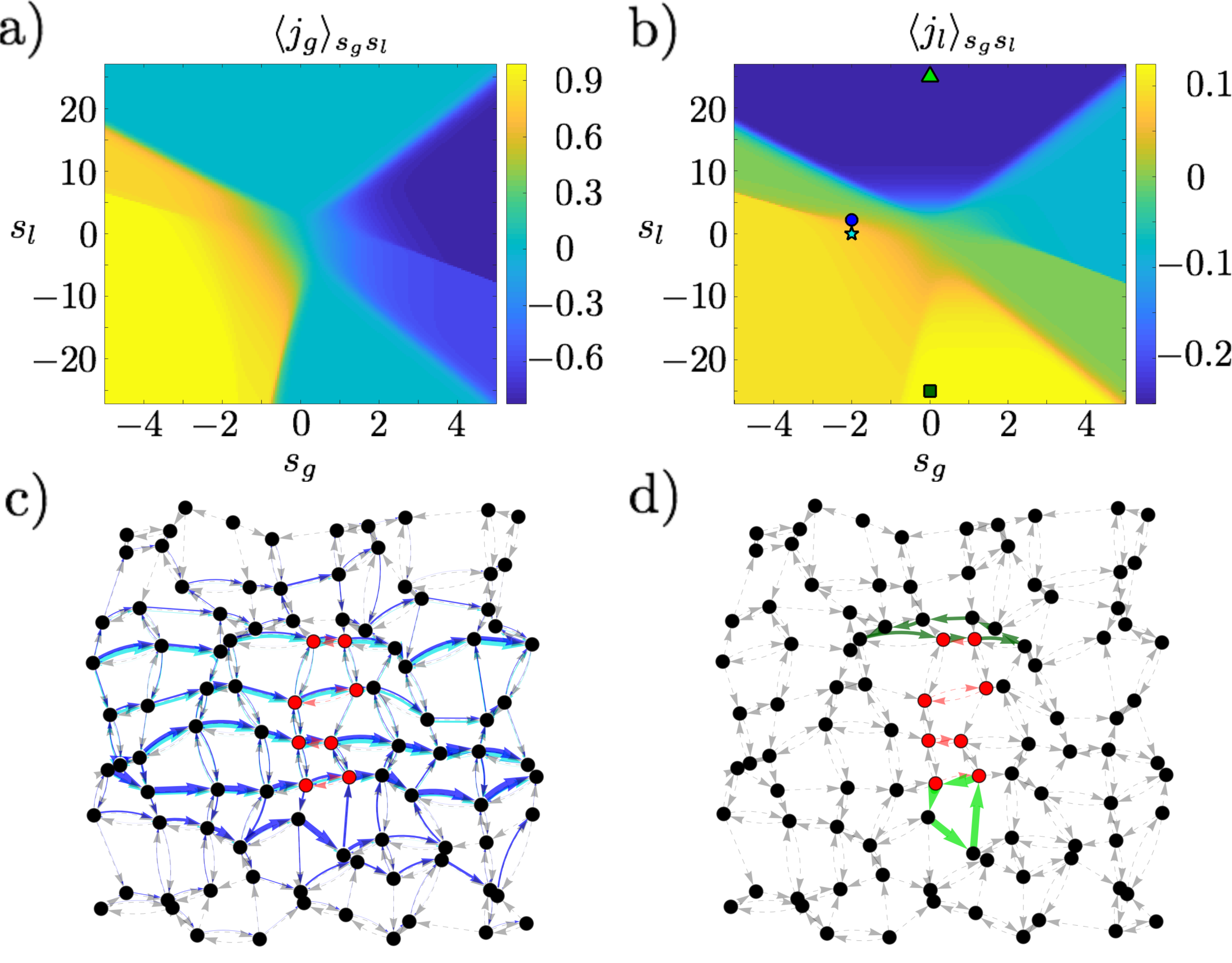}\\
\caption{{\sf \bf Optimal weight distribution for the maximization of flows under constraints in spatial networks.}
(a) Average global current $\langle j_g \rangle_{s_g s_l}$ in the graph shown in panels (c) and (d) as a function of the tilting parameters $s_g$ and $s_l$. (b) Average local current through the highlighted links joining two red nodes $\langle j_l \rangle_{s_g s_l}$ as a function of the tilting parameters $s_g$ and $s_l$. Highlighted points correspond to probability fluxes shown in the panels below. (c) Probability fluxes obtained from the Doob-transformed process corresponding to the blue circle and the cyan star shown in (b).(c) Probability fluxes obtained from the Doob-transformed process corresponding to the light green triangle and the dark green square in (b).  Probability fluxes smaller than the largest value divided by 200 are not displayed for visibility reasons.
}\label{fig3}
\end{figure}

We next illustrate how to find optimal weight distributions that maximize flows in the presence of constraints. To do so, we consider a SRW on the spatial network shown in  Fig.~\ref{fig3} (c) and (d) ---the links highlighted in colors other than gray are probability fluxes in certain dynamical regimes that will be discussed below. This network of $N=100$ nodes has been generated by adding a noise term uniformly distributed in $[-0.5, 0.5]$ to the $x$ and $y$ coordinates of the nodes of a $10\times 10$ square lattice of lattice constant one. As an additional source of disorder, the links are equiprobably set to be either bidirectional or unidirectional (in the latter case, the direction is also chosen at random). Since we consider periodic boundary conditions, these links also include those connecting one end of the network with the opposite end, both in the horizontal and vertical directions (not shown in the Figure for visibility reasons). The fact that the network is spatial is important, because the definition of the flow will depend on the coordinates of the nodes that are traversed. As in the previous section, our results are meant to be illustrative and thus we focus on a moderately large and visually simple network, but the same approach can be employed on any topology without restriction. 

Two observables are considered, a global observable and a local one. The global observable is the total current along the horizontal axis, which we denote $j_g$, and is defined as the increment  in $x$-coordinate per unit time in a random walk of $\tau$ time steps. In a jump from node $i$ to node $j$, the contribution to the current is thus $x_j - x_i$, where $x_i$ is the horizontal cartesian coordinate of node $i$. As we are interested in the flow between the leftmost and rightmost nodes of a finite spatial network, those links that (due to the periodic boundary conditions) join nodes at opposite ends in the horizontal direction do not contribute to the global current $j_g$.
%as the increment in $x$-coordinate associated with a jump between two connected nodes (in a jump from node $i$ to node $j$, the contribution to the current is $x_j - x_i$, where $x_i$ is the horizontal cartesian coordinate of node $i$). 

The local observable, on the other hand, is associated with the highlighted links joining the red nodes in the center of the graphs displayed in  Fig.~\ref{fig3} (c) and (d). We denote it $j_l$ as it is a local current, which simply adds $+1$ if the particle jumps from a red node to another red node that its lying on the right and $-1$ if the jump goes towards the left. In any other case it is zero, and that includes jumps that are predominantly in the vertical direction or which connect a red node with a black node or two black nodes. In the definition of $j_l$ we do not consider the coordinates of the red nodes, as we are just interested in limiting how frequently the walker passes through those links in a given direction. The role of predominantly horizontal links joining red nodes is in a way similar to that of the links leading to the obstacle in Fig.~\ref{fig2}, though in this case we are considering weight distributions, not optimal paths, and currents instead of activities, and both the type of random walk and the statistical ensemble are different. While $-1 < j_g < 1$, the maximum value that $j_l$ can achieve depends on the details of the network (see below). 

As both currents, $j_g$ and $j_l$, are time-averaged observables (of the form $O/\tau$, cf.\! Section II), we consider the $ss$-ensemble (see Appendix B), for which the tilted probability distribution is
\begin{equation}
P^{s_g s_l}_{\tau}(j_g,j_l) =  e^{-s_g \tau j_g - s_l \tau j _l }  P_{\tau}(j_g,j_l)/Z_{\tau}(s_g,s_l)\, ,
\end{equation}
where $ P_{\tau}(j_g,j_l)= P^{s_g=0,s_l=0}_{\tau}(j_g,j_l)$, and $Z_{\tau}(s_g,s_l)$ is a normalizing factor, and the time $\tau$ is assumed to be fixed and large. The SCGF $\theta(s_g,s_l)$ is computed as the largest eigenvalue of the corresponding tilted operator (see Appendix B), and from its partial derivatives we obtain the average global current $\langle j_g \rangle_{s_g s_l} = -\partial_{s_g} \theta(s_g,s_l)$ and the average local current $\langle j_l \rangle_{s_g s_l} = -\partial_{s_l} \theta(s_g,s_l)$, which are shown in  Fig.~\ref{fig3} (a) and (b) respectively. Several distinct regions are identified on both color maps. While for a regular lattice there is a symmetry under the simultaneous reversal of both tilting parameters $s_g \rightarrow -s_g$,  $s_l \rightarrow -s_l$, i.e.\! $\theta(-s_g,-s_l) = \theta(s_g,s_l)$, the presence of disorder breaks that symmetry in the finite disordered spatial network, leading to some quantitative differences. Such a symmetry would be recovered in the hydrodynamic limit, where microscopic structural details are coarse-grained \cite{villavicencio16a}.

We focus on global currents moving from left to right, corresponding to $s_g<0$. We first look at the case where this is the only tilting parameter ---by setting $s_l =0$---, see the $(s_g = -2, s_l =0)$ point highlighted as a cyan star in  Fig.~\ref{fig3} (b) and the probability fluxes  depicted as cyan arrows in panel (c). The very large global current $\langle j_g\rangle_{s_g s_l} \approx 0.769$ [see (a)] is due to the existence of jumps that advance consistently to the right along the fastest possible routes. It turns out that those routes go along the links joining the red nodes [see (c)], and thus the local current $\langle j_l\rangle_{s_g s_l}$ is also relatively large for this point, specifically $\langle j_l \rangle_{s_g s_l} \approx 0.078$  [see (b)]. From the Doob-transformed transition matrix used to calculate the probability fluxes, one extracts the optimal link weights that give rise to such currents, as explained in Appendix C. 

While keeping such a strong negative tilting parameter $s_g$ so as to generate a strong global current to the right, one may want to prevent the red links from being overused. In applications, this may correspond to a transportation route or communication link that may exceed its limiting capacity. To achieve that, we set $s_l$ to some positive value. Specifically, we select the value of $s_l$ that gives rise to a local current $\langle  j_l\rangle_{s_g s_l}$ that is half the value obtained for the same $s_g$ and $s_l=0$, which corresponds to the point  $(s_g = -2, s_l =2.19)$. The resulting weight distribution finds alternative routes so that the particle goes through the red links half as frequently as for $s_l=0$, while decreasing the total current by slightly less than $10 \% $. See the blue circle in panel (b) and the corresponding blue arrows in panel (c).

An interesting phenomenon occurs for very strong tilting of the local current $j_l$. See for example the points $(s_g = 0, s_l =-25)$ and $(s_g = 0, s_l =25)$, highlighted as a dark green square and a light green triangle in  Fig.~\ref{fig3} (b), respectively. They correspond to the largest and the smallest value that $\langle j_l\rangle_{s_g s_l}$ can take, which are $1/8$ and $-1/4$. These values are associated with the appearance of vortices in the trajectories, see panel (d), the absolute value of $\langle j_l \rangle_{s_g s_l}$ being the reciprocal of the number of nodes in the cycle. An analogous vortex dynamics has been found in the simple exclusion process on general graphs \cite{bodineau2008} and in the zero-range process on a diamond lattice \cite{villavicencio12a}, but its appearance in random walks with periodic boundary conditions has not been previously reported, as far as we are aware. For sufficiently small (in absolute value) $s_g$, this vortex dynamics is preserved intact, as illustrated by the blue and yellow triangle-shaped regions in (b), which shows that the tilting of the local current $s_l$ prevails over $s_g$ there. The corresponding points in panel (a) show a zero average global current, as befits such localized cyclic motion.

To conclude the exploration of the different dynamical regimes to be found over the spatial network, we focus on positive values of $s_g$, which correspond to negative global currents $j_g$. Apart from the vortex regions for very large $|s_l|$ already discussed (where the sign of $s_g$ becomes irrelevant), there is a crossover between a negative and a vanishing local current $j_l$ as $s_l$ is decreased from zero towards negative values, see Fig.~\ref{fig3} (b). This reflects the fact that, while for $s_l=0$ (or moderately positive values) the highest current is achieved by going across the links joining the red nodes, decreasing $s_l$ forces the local current to be positive, which entails avoiding those links. As a consequence, the absolute value of the global current also diminishes, see Fig.~\ref{fig3} (a). 

\section{Discussion}

We have shown how to unveil generalized optimal paths and weight distributions in networks by means of a large-deviation approach.  By combining the use of ensembles of trajectories of random walks, where time-integrated observables play the role of order parameters, and the application of the Doob transform, one can visualize the paths, or obtain the transition probabilities, that yield a certain statistical characterization of one or several observables. The statistical nature of the approach presumes that the random walk is performed numerous times, either successively in time, or by many non-interacting random walkers simultaneously. On the other hand, the results are not statistical in the sense of giving information on network ensembles (e.g.\! all random graphs with a certain degree distribution); quite on the contrary, paths and weights are found for specific topologies (i.e.\! specific realizations, if one considers the network to be an element of an ensemble), which can be arbitrary ---directed, undirected, weighted, unweighted, spatial or non-spatial. Time-dependent networks could be conceivably considered too, by including a time-dependence in the transition probabilities of the processes, but that problem is outside the scope of the present work.

The most novel aspect of our approach is that it shows how to find optimal paths and weight distributions with constraints, which do not have to be limited to one or a given number of observables. For example, one can find the shortest route between two nodes that does not pass through another one more than one fifth of the times the target node is reached, while another node is visited twice as frequently; or the optimal weighted links adapted to a certain flow without exceeding some activity threshold in some set of nodes, and another one in some specific links. Paths and weights can be tailored to a given mean value or fluctuations of essentially any time-integrated observable, so we expect the approach to be widely applicable. 

To our acknowledge, these are problems that are outside the reach of standard graph-theoretical algorithms and combinatorial approaches. Even when some proposals might exist to address one of the problems we have discussed ---which is in fact quite possible given the large literature on the subject, spanning various fields of science and engineering---, shortest-path and related algorithms are typically specific to the problem at hand, small qualitative changes in the constraints requiring important methodological modifications, while our statistical physics approach is very flexible in this regard. The examples we have shown are meant to be simple and illustrative, but more complex scenarios (in terms of network topology, observables and constraints) can be similarly studied. All depends on a judicious choice of the appropriate process ---which type of random walk, though processes involving exclusion or others could also be considered---, observables and statistical ensemble. 

 The computational complexity of this approach will be that of the algorithm used to extract the largest eigenvalue (whose logarithm is the SCGF) and the associated eigenvectors of the tilted generator. While the latter is typically a sparse matrix, which may constitute a significant numerical advantage, for very large networks the numerical eigenvalue problem may be challenging. In such situations, the large-deviation function can be obtained from numerical approaches based on the cloning algorithm \cite{giardina06a,giardina11a,carollo20a} or adaptive sampling \cite{nemoto2017,ferre18a}. Moreover, the intriguing possibility of finding the optimal dynamics leading to a prescribed fluctuation by approaches adapted from reinforcement learning has emerged lately \cite{rose20a}. Similar machine-learning formulations have been recently proposed for dealing with numerically intractable optimization problems in condensed-matter physics \cite{whitelam20a,barr20a}.

\begin{acknowledgments}
We thank Juan P. Garrahan, from whom we have learned much about the thermodynamics-of-trajectories formalism used in this paper, for useful suggestions. We are also grateful to H. Touchette for reading the manuscript and for helpful comments. C.P.E. acknowledges C. Giardin\`a and C. Giberti for insightful discussions. The research leading to these results has received funding from the European Union's Horizon 2020 research and innovation programme under the Marie Sklodowska-Curie Cofund Programme Athenea3I Grant Agreement No. 754446, from the European Regional Development Fund, Junta de Andaluc\'ia-Consejer\'ia de Econom\'ia y Conocimiento, Ref. A-FQM-175-UGR18, and from MINECO, Spain (FIS2017-84151-P).  We are grateful for the computing resources and related technical support provided by PROTEUS, the super-computing center of Institute Carlos I in Granada, Spain, and by CRESCO/ENEAGRID High Performance Computing infrastructure and its staff \cite{iannone2019}, which is funded by ENEA, the Italian National Agency for New Technologies, Energy and Sustainable Economic Development and by Italian and European research programmes.
\end{acknowledgments}

\bibliography{OptPathsLDBib}{}

\begin{thebibliography}{68}%
\makeatletter
\providecommand \@ifxundefined [1]{%
 \@ifx{#1\undefined}
}%
\providecommand \@ifnum [1]{%
 \ifnum #1\expandafter \@firstoftwo
 \else \expandafter \@secondoftwo
 \fi
}%
\providecommand \@ifx [1]{%
 \ifx #1\expandafter \@firstoftwo
 \else \expandafter \@secondoftwo
 \fi
}%
\providecommand \natexlab [1]{#1}%
\providecommand \enquote  [1]{``#1''}%
\providecommand \bibnamefont  [1]{#1}%
\providecommand \bibfnamefont [1]{#1}%
\providecommand \citenamefont [1]{#1}%
\providecommand \href@noop [0]{\@secondoftwo}%
\providecommand \href [0]{\begingroup \@sanitize@url \@href}%
\providecommand \@href[1]{\@@startlink{#1}\@@href}%
\providecommand \@@href[1]{\endgroup#1\@@endlink}%
\providecommand \@sanitize@url [0]{\catcode `\\12\catcode `\$12\catcode
  `\&12\catcode `\#12\catcode `\^12\catcode `\_12\catcode `\%12\relax}%
\providecommand \@@startlink[1]{}%
\providecommand \@@endlink[0]{}%
\providecommand \url  [0]{\begingroup\@sanitize@url \@url }%
\providecommand \@url [1]{\endgroup\@href {#1}{\urlprefix }}%
\providecommand \urlprefix  [0]{URL }%
\providecommand \Eprint [0]{\href }%
\providecommand \doibase [0]{http://dx.doi.org/}%
\providecommand \selectlanguage [0]{\@gobble}%
\providecommand \bibinfo  [0]{\@secondoftwo}%
\providecommand \bibfield  [0]{\@secondoftwo}%
\providecommand \translation [1]{[#1]}%
\providecommand \BibitemOpen [0]{}%
\providecommand \bibitemStop [0]{}%
\providecommand \bibitemNoStop [0]{.\EOS\space}%
\providecommand \EOS [0]{\spacefactor3000\relax}%
\providecommand \BibitemShut  [1]{\csname bibitem#1\endcsname}%
\let\auto@bib@innerbib\@empty
%</preamble>
\bibitem [{\citenamefont {Zhan}\ and\ \citenamefont {Noon}(1998)}]{zhan1998}%
  \BibitemOpen
  \bibfield  {author} {\bibinfo {author} {\bibfnamefont {F.~B.}\ \bibnamefont
  {Zhan}}\ and\ \bibinfo {author} {\bibfnamefont {C.~E.}\ \bibnamefont
  {Noon}},\ }\bibfield  {title} {\enquote {\bibinfo {title} {Shortest path
  algorithms: an evaluation using real road networks},}\ }\href@noop {}
  {\bibfield  {journal} {\bibinfo  {journal} {Transp. Sci.}\ }\textbf {\bibinfo
  {volume} {32}},\ \bibinfo {pages} {65} (\bibinfo {year} {1998})}\BibitemShut
  {NoStop}%
\bibitem [{\citenamefont {Zeng}\ and\ \citenamefont {Church}(2009)}]{zeng2009}%
  \BibitemOpen
  \bibfield  {author} {\bibinfo {author} {\bibfnamefont {W.}~\bibnamefont
  {Zeng}}\ and\ \bibinfo {author} {\bibfnamefont {R.~L.}\ \bibnamefont
  {Church}},\ }\bibfield  {title} {\enquote {\bibinfo {title} {Finding shortest
  paths on real road networks: the case for {A}*},}\ }\href@noop {} {\bibfield
  {journal} {\bibinfo  {journal} {Int. J. Geogr. Inf. Sci}\ }\textbf {\bibinfo
  {volume} {23}},\ \bibinfo {pages} {531} (\bibinfo {year} {2009})}\BibitemShut
  {NoStop}%
\bibitem [{\citenamefont {Li}\ \emph {et~al.}(2010)\citenamefont {Li},
  \citenamefont {Reis}, \citenamefont {Moreira}, \citenamefont {Havlin},
  \citenamefont {Stanley},\ and\ \citenamefont {Andrade~Jr.}}]{li2010}%
  \BibitemOpen
  \bibfield  {author} {\bibinfo {author} {\bibfnamefont {G.}~\bibnamefont
  {Li}}, \bibinfo {author} {\bibfnamefont {S.~D.~S.}\ \bibnamefont {Reis}},
  \bibinfo {author} {\bibfnamefont {A.~A.}\ \bibnamefont {Moreira}}, \bibinfo
  {author} {\bibfnamefont {S.}~\bibnamefont {Havlin}}, \bibinfo {author}
  {\bibfnamefont {H.~E.}\ \bibnamefont {Stanley}}, \ and\ \bibinfo {author}
  {\bibfnamefont {J.~S.}\ \bibnamefont {Andrade~Jr.}},\ }\bibfield  {title}
  {\enquote {\bibinfo {title} {Towards design principles for optimal transport
  networks},}\ }\href@noop {} {\bibfield  {journal} {\bibinfo  {journal} {Phys.
  Rev. Lett.}\ }\textbf {\bibinfo {volume} {104}},\ \bibinfo {pages} {018701}
  (\bibinfo {year} {2010})}\BibitemShut {NoStop}%
\bibitem [{\citenamefont {Albert}\ \emph {et~al.}(1999)\citenamefont {Albert},
  \citenamefont {Jeong},\ and\ \citenamefont {Barab{\'a}si}}]{albert1999}%
  \BibitemOpen
  \bibfield  {author} {\bibinfo {author} {\bibfnamefont {R.}~\bibnamefont
  {Albert}}, \bibinfo {author} {\bibfnamefont {H.}~\bibnamefont {Jeong}}, \
  and\ \bibinfo {author} {\bibfnamefont {A.-L.}\ \bibnamefont {Barab{\'a}si}},\
  }\bibfield  {title} {\enquote {\bibinfo {title} {Diameter of the world-wide
  web},}\ }\href@noop {} {\bibfield  {journal} {\bibinfo  {journal} {Nature}\
  }\textbf {\bibinfo {volume} {401}},\ \bibinfo {pages} {130} (\bibinfo {year}
  {1999})}\BibitemShut {NoStop}%
\bibitem [{\citenamefont {{Tangmunarunkit}}\ \emph {et~al.}(2001)\citenamefont
  {{Tangmunarunkit}}, \citenamefont {{Govindan}}, \citenamefont {{Shenker}},\
  and\ \citenamefont {{Estrin}}}]{tangmunarunkit2001}%
  \BibitemOpen
  \bibfield  {author} {\bibinfo {author} {\bibfnamefont {H.}~\bibnamefont
  {{Tangmunarunkit}}}, \bibinfo {author} {\bibfnamefont {R.}~\bibnamefont
  {{Govindan}}}, \bibinfo {author} {\bibfnamefont {S.}~\bibnamefont
  {{Shenker}}}, \ and\ \bibinfo {author} {\bibfnamefont {D.}~\bibnamefont
  {{Estrin}}},\ }\bibfield  {title} {\enquote {\bibinfo {title} {The impact of
  routing policy on internet paths},}\ }in\ \href@noop {} {\emph {\bibinfo
  {booktitle} {Proceedings IEEE INFOCOM 2001. Conference on Computer
  Communications. Twentieth Annual Joint Conference of the IEEE Computer and
  Communications Society (Cat. No.01CH37213)}}},\ Vol.~\bibinfo {volume} {2}\
  (\bibinfo {year} {2001})\ pp.\ \bibinfo {pages} {736--742 vol.2}\BibitemShut
  {NoStop}%
\bibitem [{\citenamefont {Fortz}\ and\ \citenamefont
  {Thorup}(2004)}]{fortz2004}%
  \BibitemOpen
  \bibfield  {author} {\bibinfo {author} {\bibfnamefont {B.}~\bibnamefont
  {Fortz}}\ and\ \bibinfo {author} {\bibfnamefont {M.}~\bibnamefont {Thorup}},\
  }\bibfield  {title} {\enquote {\bibinfo {title} {Increasing internet capacity
  using local search},}\ }\href@noop {} {\bibfield  {journal} {\bibinfo
  {journal} {Comput. Optim. Appl.}\ }\textbf {\bibinfo {volume} {29}},\
  \bibinfo {pages} {13} (\bibinfo {year} {2004})}\BibitemShut {NoStop}%
\bibitem [{\citenamefont {Echenique}\ \emph {et~al.}(2004)\citenamefont
  {Echenique}, \citenamefont {G{\'o}mez-Garde{\~n}es},\ and\ \citenamefont
  {Moreno}}]{echenique2004}%
  \BibitemOpen
  \bibfield  {author} {\bibinfo {author} {\bibfnamefont {P.}~\bibnamefont
  {Echenique}}, \bibinfo {author} {\bibfnamefont {J.}~\bibnamefont
  {G{\'o}mez-Garde{\~n}es}}, \ and\ \bibinfo {author} {\bibfnamefont
  {Y.}~\bibnamefont {Moreno}},\ }\bibfield  {title} {\enquote {\bibinfo {title}
  {Improved routing strategies for internet traffic delivery},}\ }\href@noop {}
  {\bibfield  {journal} {\bibinfo  {journal} {Phys. Rev. E}\ }\textbf {\bibinfo
  {volume} {70}},\ \bibinfo {pages} {056105} (\bibinfo {year}
  {2004})}\BibitemShut {NoStop}%
\bibitem [{\citenamefont {Yan}\ \emph {et~al.}(2006)\citenamefont {Yan},
  \citenamefont {Zhou}, \citenamefont {Hu}, \citenamefont {Fu},\ and\
  \citenamefont {Wang}}]{yan2006}%
  \BibitemOpen
  \bibfield  {author} {\bibinfo {author} {\bibfnamefont {G.}~\bibnamefont
  {Yan}}, \bibinfo {author} {\bibfnamefont {T.}~\bibnamefont {Zhou}}, \bibinfo
  {author} {\bibfnamefont {B.}~\bibnamefont {Hu}}, \bibinfo {author}
  {\bibfnamefont {Z.-Q.}\ \bibnamefont {Fu}}, \ and\ \bibinfo {author}
  {\bibfnamefont {B.-H.}\ \bibnamefont {Wang}},\ }\bibfield  {title} {\enquote
  {\bibinfo {title} {Efficient routing on complex networks},}\ }\href@noop {}
  {\bibfield  {journal} {\bibinfo  {journal} {Phys. Rev. E}\ }\textbf {\bibinfo
  {volume} {73}},\ \bibinfo {pages} {046108} (\bibinfo {year}
  {2006})}\BibitemShut {NoStop}%
\bibitem [{\citenamefont {Managbanag}\ \emph {et~al.}(2008)\citenamefont
  {Managbanag}, \citenamefont {Witten}, \citenamefont {Bonchev}, \citenamefont
  {Fox}, \citenamefont {Tsuchiya}, \citenamefont {Kennedy},\ and\ \citenamefont
  {Kaeberlein}}]{managbanag2008}%
  \BibitemOpen
  \bibfield  {author} {\bibinfo {author} {\bibfnamefont {J.~R.}\ \bibnamefont
  {Managbanag}}, \bibinfo {author} {\bibfnamefont {T.~M.}\ \bibnamefont
  {Witten}}, \bibinfo {author} {\bibfnamefont {D.}~\bibnamefont {Bonchev}},
  \bibinfo {author} {\bibfnamefont {L.~A.}\ \bibnamefont {Fox}}, \bibinfo
  {author} {\bibfnamefont {M.}~\bibnamefont {Tsuchiya}}, \bibinfo {author}
  {\bibfnamefont {B.~K.}\ \bibnamefont {Kennedy}}, \ and\ \bibinfo {author}
  {\bibfnamefont {M.}~\bibnamefont {Kaeberlein}},\ }\bibfield  {title}
  {\enquote {\bibinfo {title} {Shortest-path network analysis is a useful
  approach toward identifying genetic determinants of longevity},}\ }\href@noop
  {} {\bibfield  {journal} {\bibinfo  {journal} {PLOS ONE}\ }\textbf {\bibinfo
  {volume} {3}},\ \bibinfo {pages} {e3802} (\bibinfo {year}
  {2008})}\BibitemShut {NoStop}%
\bibitem [{\citenamefont {Jiang}\ \emph {et~al.}(2013)\citenamefont {Jiang},
  \citenamefont {Chen}, \citenamefont {Zhang}, \citenamefont {Chen},
  \citenamefont {Zhang}, \citenamefont {Huang}, \citenamefont {Cai},\ and\
  \citenamefont {Kong}}]{jiang2013}%
  \BibitemOpen
  \bibfield  {author} {\bibinfo {author} {\bibfnamefont {M.}~\bibnamefont
  {Jiang}}, \bibinfo {author} {\bibfnamefont {Y.}~\bibnamefont {Chen}},
  \bibinfo {author} {\bibfnamefont {Y.}~\bibnamefont {Zhang}}, \bibinfo
  {author} {\bibfnamefont {L.}~\bibnamefont {Chen}}, \bibinfo {author}
  {\bibfnamefont {N.}~\bibnamefont {Zhang}}, \bibinfo {author} {\bibfnamefont
  {T.}~\bibnamefont {Huang}}, \bibinfo {author} {\bibfnamefont {Y.-D.}\
  \bibnamefont {Cai}}, \ and\ \bibinfo {author} {\bibfnamefont {X.-Y.}\
  \bibnamefont {Kong}},\ }\bibfield  {title} {\enquote {\bibinfo {title}
  {Identification of hepatocellular carcinoma related genes with k-th shortest
  paths in a protein--protein interaction network},}\ }\href@noop {} {\bibfield
   {journal} {\bibinfo  {journal} {Mol. BioSyst.}\ }\textbf {\bibinfo {volume}
  {9}},\ \bibinfo {pages} {2720} (\bibinfo {year} {2013})}\BibitemShut
  {NoStop}%
\bibitem [{\citenamefont {Boccaletti}\ \emph {et~al.}(2006)\citenamefont
  {Boccaletti}, \citenamefont {Latora}, \citenamefont {Moreno}, \citenamefont
  {Chavez},\ and\ \citenamefont {Hwang}}]{boccaletti2006}%
  \BibitemOpen
  \bibfield  {author} {\bibinfo {author} {\bibfnamefont {S.}~\bibnamefont
  {Boccaletti}}, \bibinfo {author} {\bibfnamefont {V.}~\bibnamefont {Latora}},
  \bibinfo {author} {\bibfnamefont {Y.}~\bibnamefont {Moreno}}, \bibinfo
  {author} {\bibfnamefont {M.}~\bibnamefont {Chavez}}, \ and\ \bibinfo {author}
  {\bibfnamefont {D.-U.}\ \bibnamefont {Hwang}},\ }\bibfield  {title} {\enquote
  {\bibinfo {title} {Complex networks: {S}tructure and dynamics},}\ }\href@noop
  {} {\bibfield  {journal} {\bibinfo  {journal} {Phys. Rep.}\ }\textbf
  {\bibinfo {volume} {424}},\ \bibinfo {pages} {175} (\bibinfo {year}
  {2006})}\BibitemShut {NoStop}%
\bibitem [{\citenamefont {Newman}(2010)}]{newman2010}%
  \BibitemOpen
  \bibfield  {author} {\bibinfo {author} {\bibfnamefont {M.~E.~J.}\
  \bibnamefont {Newman}},\ }\href@noop {} {\emph {\bibinfo {title} {Networks:
  {A}n {I}ntroduction}}}\ (\bibinfo  {publisher} {Oxford University Press},\
  \bibinfo {address} {Oxford},\ \bibinfo {year} {2010})\BibitemShut {NoStop}%
\bibitem [{\citenamefont {Zheng}\ \emph {et~al.}(1996)\citenamefont {Zheng},
  \citenamefont {Lim},\ and\ \citenamefont {Iyengar}}]{zheng1996}%
  \BibitemOpen
  \bibfield  {author} {\bibinfo {author} {\bibfnamefont {S.-Q.}\ \bibnamefont
  {Zheng}}, \bibinfo {author} {\bibfnamefont {J.~S.}\ \bibnamefont {Lim}}, \
  and\ \bibinfo {author} {\bibfnamefont {S.~S.}\ \bibnamefont {Iyengar}},\
  }\bibfield  {title} {\enquote {\bibinfo {title} {Finding obstacle-avoiding
  shortest paths using implicit connection graphs},}\ }\href@noop {} {\bibfield
   {journal} {\bibinfo  {journal} {IEEE Trans. Comput.-Aided Des. Integr.
  Circuits Syst.}\ }\textbf {\bibinfo {volume} {15}},\ \bibinfo {pages} {103}
  (\bibinfo {year} {1996})}\BibitemShut {NoStop}%
\bibitem [{\citenamefont {Carlyle}\ \emph {et~al.}(2008)\citenamefont
  {Carlyle}, \citenamefont {Royset},\ and\ \citenamefont {Wood}}]{carlyle2008}%
  \BibitemOpen
  \bibfield  {author} {\bibinfo {author} {\bibfnamefont {W.~M.}\ \bibnamefont
  {Carlyle}}, \bibinfo {author} {\bibfnamefont {J.~O}\ \bibnamefont {Royset}},
  \ and\ \bibinfo {author} {\bibfnamefont {R.~K.}\ \bibnamefont {Wood}},\
  }\bibfield  {title} {\enquote {\bibinfo {title} {Lagrangian relaxation and
  enumeration for solving constrained shortest-path problems},}\ }\href@noop {}
  {\bibfield  {journal} {\bibinfo  {journal} {Networks}\ }\textbf {\bibinfo
  {volume} {52}},\ \bibinfo {pages} {256} (\bibinfo {year} {2008})}\BibitemShut
  {NoStop}%
\bibitem [{\citenamefont {Lozano}\ and\ \citenamefont
  {Medaglia}(2013)}]{lozano2013}%
  \BibitemOpen
  \bibfield  {author} {\bibinfo {author} {\bibfnamefont {L.}~\bibnamefont
  {Lozano}}\ and\ \bibinfo {author} {\bibfnamefont {A.~L.}\ \bibnamefont
  {Medaglia}},\ }\bibfield  {title} {\enquote {\bibinfo {title} {On an exact
  method for the constrained shortest path problem},}\ }\href@noop {}
  {\bibfield  {journal} {\bibinfo  {journal} {Comput. Oper. Res.}\ }\textbf
  {\bibinfo {volume} {40}},\ \bibinfo {pages} {378} (\bibinfo {year}
  {2013})}\BibitemShut {NoStop}%
\bibitem [{\citenamefont {Hershberger}\ \emph {et~al.}(2020)\citenamefont
  {Hershberger}, \citenamefont {Kumar},\ and\ \citenamefont
  {Suri}}]{hershberger2020}%
  \BibitemOpen
  \bibfield  {author} {\bibinfo {author} {\bibfnamefont {J.}~\bibnamefont
  {Hershberger}}, \bibinfo {author} {\bibfnamefont {N.}~\bibnamefont {Kumar}},
  \ and\ \bibinfo {author} {\bibfnamefont {S.}~\bibnamefont {Suri}},\
  }\bibfield  {title} {\enquote {\bibinfo {title} {Shortest paths in the plane
  with obstacle violations},}\ }\href@noop {} {\bibfield  {journal} {\bibinfo
  {journal} {Algorithmica}\ }\textbf {\bibinfo {volume} {82}},\ \bibinfo
  {pages} {1813} (\bibinfo {year} {2020})}\BibitemShut {NoStop}%
\bibitem [{\citenamefont {Knopp}\ \emph {et~al.}(2007)\citenamefont {Knopp},
  \citenamefont {Sanders}, \citenamefont {Schultes}, \citenamefont {Schulz},\
  and\ \citenamefont {Wagner}}]{knopp2007}%
  \BibitemOpen
  \bibfield  {author} {\bibinfo {author} {\bibfnamefont {S.}~\bibnamefont
  {Knopp}}, \bibinfo {author} {\bibfnamefont {P.}~\bibnamefont {Sanders}},
  \bibinfo {author} {\bibfnamefont {D.}~\bibnamefont {Schultes}}, \bibinfo
  {author} {\bibfnamefont {F.}~\bibnamefont {Schulz}}, \ and\ \bibinfo {author}
  {\bibfnamefont {D.}~\bibnamefont {Wagner}},\ }\bibfield  {title} {\enquote
  {\bibinfo {title} {Computing many-to-many shortest paths using highway
  hierarchies},}\ }in\ \href@noop {} {\emph {\bibinfo {booktitle} {2007
  Proceedings of the Ninth Workshop on Algorithm Engineering and Experiments
  (ALENEX)}}}\ (\bibinfo {organization} {SIAM},\ \bibinfo {year} {2007})\
  p.~\bibinfo {pages} {36}\BibitemShut {NoStop}%
\bibitem [{\citenamefont {Jagadeesh}\ and\ \citenamefont
  {Srikanthan}(2019)}]{jagadeesh2019}%
  \BibitemOpen
  \bibfield  {author} {\bibinfo {author} {\bibfnamefont {G.~R.}\ \bibnamefont
  {Jagadeesh}}\ and\ \bibinfo {author} {\bibfnamefont {T.}~\bibnamefont
  {Srikanthan}},\ }\bibfield  {title} {\enquote {\bibinfo {title} {Fast
  computation of clustered many-to-many shortest paths and its application to
  map matching},}\ }\href@noop {} {\bibfield  {journal} {\bibinfo  {journal}
  {ACM Trans. Spat. Algorithms Syst.}\ }\textbf {\bibinfo {volume} {5}},\
  \bibinfo {pages} {1} (\bibinfo {year} {2019})}\BibitemShut {NoStop}%
\bibitem [{\citenamefont {Tanaka}\ and\ \citenamefont
  {Aoyagi}(2008)}]{tanaka2008}%
  \BibitemOpen
  \bibfield  {author} {\bibinfo {author} {\bibfnamefont {T.}~\bibnamefont
  {Tanaka}}\ and\ \bibinfo {author} {\bibfnamefont {T.}~\bibnamefont
  {Aoyagi}},\ }\bibfield  {title} {\enquote {\bibinfo {title} {Optimal weighted
  networks of phase oscillators for synchronization},}\ }\href@noop {}
  {\bibfield  {journal} {\bibinfo  {journal} {Phys. Rev. E}\ }\textbf {\bibinfo
  {volume} {78}},\ \bibinfo {pages} {046210} (\bibinfo {year}
  {2008})}\BibitemShut {NoStop}%
\bibitem [{\citenamefont {Nishikawa}\ and\ \citenamefont
  {Motter}(2006)}]{nishikawa2006}%
  \BibitemOpen
  \bibfield  {author} {\bibinfo {author} {\bibfnamefont {T.}~\bibnamefont
  {Nishikawa}}\ and\ \bibinfo {author} {\bibfnamefont {A.~E.}\ \bibnamefont
  {Motter}},\ }\bibfield  {title} {\enquote {\bibinfo {title} {Synchronization
  is optimal in nondiagonalizable networks},}\ }\href@noop {} {\bibfield
  {journal} {\bibinfo  {journal} {Phys. Rev. E}\ }\textbf {\bibinfo {volume}
  {73}},\ \bibinfo {pages} {065106(R)} (\bibinfo {year} {2006})}\BibitemShut
  {NoStop}%
\bibitem [{\citenamefont {Sevilla-Escoboza}\ \emph {et~al.}(2016)\citenamefont
  {Sevilla-Escoboza}, \citenamefont {Buld{\'u}}, \citenamefont {Boccaletti},
  \citenamefont {Papo}, \citenamefont {Hwang}, \citenamefont {Huerta-Cuellar},\
  and\ \citenamefont {Guti{\'e}rrez}}]{sevilla2016}%
  \BibitemOpen
  \bibfield  {author} {\bibinfo {author} {\bibfnamefont {R.}~\bibnamefont
  {Sevilla-Escoboza}}, \bibinfo {author} {\bibfnamefont {J.~M.}\ \bibnamefont
  {Buld{\'u}}}, \bibinfo {author} {\bibfnamefont {S.}~\bibnamefont
  {Boccaletti}}, \bibinfo {author} {\bibfnamefont {D.}~\bibnamefont {Papo}},
  \bibinfo {author} {\bibfnamefont {D.-U.}\ \bibnamefont {Hwang}}, \bibinfo
  {author} {\bibfnamefont {G.}~\bibnamefont {Huerta-Cuellar}}, \ and\ \bibinfo
  {author} {\bibfnamefont {R.}~\bibnamefont {Guti{\'e}rrez}},\ }\bibfield
  {title} {\enquote {\bibinfo {title} {Experimental implementation of maximally
  synchronizable networks},}\ }\href@noop {} {\bibfield  {journal} {\bibinfo
  {journal} {Physica A}\ }\textbf {\bibinfo {volume} {100}},\ \bibinfo {pages}
  {113} (\bibinfo {year} {2016})}\BibitemShut {NoStop}%
\bibitem [{\citenamefont {Kempton}\ \emph {et~al.}(2017)\citenamefont
  {Kempton}, \citenamefont {Herrmann},\ and\ \citenamefont
  {Di~Bernardo}}]{kempton2017}%
  \BibitemOpen
  \bibfield  {author} {\bibinfo {author} {\bibfnamefont {L.}~\bibnamefont
  {Kempton}}, \bibinfo {author} {\bibfnamefont {G.}~\bibnamefont {Herrmann}}, \
  and\ \bibinfo {author} {\bibfnamefont {M.}~\bibnamefont {Di~Bernardo}},\
  }\bibfield  {title} {\enquote {\bibinfo {title} {Self-organization of
  weighted networks for optimal synchronizability},}\ }\href@noop {} {\bibfield
   {journal} {\bibinfo  {journal} {IEEE Trans. Control. Netw. Syst.}\ }\textbf
  {\bibinfo {volume} {5}},\ \bibinfo {pages} {1541} (\bibinfo {year}
  {2017})}\BibitemShut {NoStop}%
\bibitem [{\citenamefont {Nemoto}\ \emph {et~al.}(2017)\citenamefont {Nemoto},
  \citenamefont {Jack},\ and\ \citenamefont {Lecomte}}]{nemoto2017}%
  \BibitemOpen
  \bibfield  {author} {\bibinfo {author} {\bibfnamefont {T.}~\bibnamefont
  {Nemoto}}, \bibinfo {author} {\bibfnamefont {R.~L.}\ \bibnamefont {Jack}}, \
  and\ \bibinfo {author} {\bibfnamefont {V.}~\bibnamefont {Lecomte}},\
  }\bibfield  {title} {\enquote {\bibinfo {title} {Finite-size scaling of a
  first-order dynamical phase transition: {A}daptive population dynamics and an
  effective model},}\ }\href@noop {} {\bibfield  {journal} {\bibinfo  {journal}
  {Phys. Rev. Lett.}\ }\textbf {\bibinfo {volume} {118}},\ \bibinfo {pages}
  {115702} (\bibinfo {year} {2017})}\BibitemShut {NoStop}%
\bibitem [{\citenamefont {Ray}\ \emph {et~al.}(2018)\citenamefont {Ray},
  \citenamefont {Chan},\ and\ \citenamefont {Limmer}}]{ray18a}%
  \BibitemOpen
  \bibfield  {author} {\bibinfo {author} {\bibfnamefont {U.}~\bibnamefont
  {Ray}}, \bibinfo {author} {\bibfnamefont {G.~K.-L.}\ \bibnamefont {Chan}}, \
  and\ \bibinfo {author} {\bibfnamefont {D.~T.}\ \bibnamefont {Limmer}},\
  }\bibfield  {title} {\enquote {\bibinfo {title} {Exact fluctuations of
  nonequilibrium steady states from approximate auxiliary dynamics},}\
  }\href@noop {} {\bibfield  {journal} {\bibinfo  {journal} {Phys. Rev. Lett.}\
  }\textbf {\bibinfo {volume} {120}},\ \bibinfo {pages} {210602} (\bibinfo
  {year} {2018})}\BibitemShut {NoStop}%
\bibitem [{\citenamefont {Hurtado-Guti{\'e}rrez}\ \emph
  {et~al.}(2020)\citenamefont {Hurtado-Guti{\'e}rrez}, \citenamefont {Carollo},
  \citenamefont {P{\'e}rez-Espigares},\ and\ \citenamefont
  {Hurtado}}]{hurtadogutierrez2020}%
  \BibitemOpen
  \bibfield  {author} {\bibinfo {author} {\bibfnamefont {R.}~\bibnamefont
  {Hurtado-Guti{\'e}rrez}}, \bibinfo {author} {\bibfnamefont {F.}~\bibnamefont
  {Carollo}}, \bibinfo {author} {\bibfnamefont {C.}~\bibnamefont
  {P{\'e}rez-Espigares}}, \ and\ \bibinfo {author} {\bibfnamefont {P.~I.}\
  \bibnamefont {Hurtado}},\ }\bibfield  {title} {\enquote {\bibinfo {title}
  {Building continuous time crystals from rare events},}\ }\href@noop {}
  {\bibfield  {journal} {\bibinfo  {journal} {Phys. Rev. Lett.}\ }\textbf
  {\bibinfo {volume} {125}},\ \bibinfo {pages} {160601} (\bibinfo {year}
  {2020})}\BibitemShut {NoStop}%
\bibitem [{\citenamefont {Lecomte}\ \emph {et~al.}(2007)\citenamefont
  {Lecomte}, \citenamefont {Appert-Rolland},\ and\ \citenamefont
  {Van~Wijland}}]{lecomte2007}%
  \BibitemOpen
  \bibfield  {author} {\bibinfo {author} {\bibfnamefont {V.}~\bibnamefont
  {Lecomte}}, \bibinfo {author} {\bibfnamefont {C.}~\bibnamefont
  {Appert-Rolland}}, \ and\ \bibinfo {author} {\bibfnamefont {F.}~\bibnamefont
  {Van~Wijland}},\ }\bibfield  {title} {\enquote {\bibinfo {title}
  {Thermodynamic formalism for systems with {M}arkov dynamics},}\ }\href@noop
  {} {\bibfield  {journal} {\bibinfo  {journal} {J. Stat. Phys.}\ }\textbf
  {\bibinfo {volume} {127}},\ \bibinfo {pages} {51} (\bibinfo {year}
  {2007})}\BibitemShut {NoStop}%
\bibitem [{\citenamefont {Garrahan}\ \emph {et~al.}(2009)\citenamefont
  {Garrahan}, \citenamefont {Jack}, \citenamefont {Lecomte}, \citenamefont
  {Pitard}, \citenamefont {van Duijvendijk},\ and\ \citenamefont {van
  Wijland}}]{garrahan2009}%
  \BibitemOpen
  \bibfield  {author} {\bibinfo {author} {\bibfnamefont {J.~P.}\ \bibnamefont
  {Garrahan}}, \bibinfo {author} {\bibfnamefont {R.~L.}\ \bibnamefont {Jack}},
  \bibinfo {author} {\bibfnamefont {V.}~\bibnamefont {Lecomte}}, \bibinfo
  {author} {\bibfnamefont {E.}~\bibnamefont {Pitard}}, \bibinfo {author}
  {\bibfnamefont {K.}~\bibnamefont {van Duijvendijk}}, \ and\ \bibinfo {author}
  {\bibfnamefont {F.}~\bibnamefont {van Wijland}},\ }\bibfield  {title}
  {\enquote {\bibinfo {title} {First-order dynamical phase transition in models
  of glasses: an approach based on ensembles of histories},}\ }\href@noop {}
  {\bibfield  {journal} {\bibinfo  {journal} {J. Phys. A: Math. Theor.}\
  }\textbf {\bibinfo {volume} {42}},\ \bibinfo {pages} {075007} (\bibinfo
  {year} {2009})}\BibitemShut {NoStop}%
\bibitem [{\citenamefont {Touchette}(2009)}]{touchette2009}%
  \BibitemOpen
  \bibfield  {author} {\bibinfo {author} {\bibfnamefont {H.}~\bibnamefont
  {Touchette}},\ }\bibfield  {title} {\enquote {\bibinfo {title} {The large
  deviation approach to statistical mechanics},}\ }\href@noop {} {\bibfield
  {journal} {\bibinfo  {journal} {Phys. Rep.}\ }\textbf {\bibinfo {volume}
  {478}},\ \bibinfo {pages} {1} (\bibinfo {year} {2009})}\BibitemShut {NoStop}%
\bibitem [{\citenamefont {De~Bacco}\ \emph {et~al.}(2016)\citenamefont
  {De~Bacco}, \citenamefont {Guggiola}, \citenamefont {K{\"u}hn},\ and\
  \citenamefont {Paga}}]{debacco2016}%
  \BibitemOpen
  \bibfield  {author} {\bibinfo {author} {\bibfnamefont {C.}~\bibnamefont
  {De~Bacco}}, \bibinfo {author} {\bibfnamefont {A.}~\bibnamefont {Guggiola}},
  \bibinfo {author} {\bibfnamefont {R.}~\bibnamefont {K{\"u}hn}}, \ and\
  \bibinfo {author} {\bibfnamefont {P.}~\bibnamefont {Paga}},\ }\bibfield
  {title} {\enquote {\bibinfo {title} {Rare events statistics of random walks
  on networks: localisation and other dynamical phase transitions},}\
  }\href@noop {} {\bibfield  {journal} {\bibinfo  {journal} {J. Phys. A: Math.
  Theor.}\ }\textbf {\bibinfo {volume} {49}},\ \bibinfo {pages} {184003}
  (\bibinfo {year} {2016})}\BibitemShut {NoStop}%
\bibitem [{\citenamefont {Coghi}\ \emph {et~al.}(2019)\citenamefont {Coghi},
  \citenamefont {Morand},\ and\ \citenamefont {Touchette}}]{coghi2019}%
  \BibitemOpen
  \bibfield  {author} {\bibinfo {author} {\bibfnamefont {F.}~\bibnamefont
  {Coghi}}, \bibinfo {author} {\bibfnamefont {J.}~\bibnamefont {Morand}}, \
  and\ \bibinfo {author} {\bibfnamefont {H.}~\bibnamefont {Touchette}},\
  }\bibfield  {title} {\enquote {\bibinfo {title} {Large deviations of random
  walks on random graphs},}\ }\href@noop {} {\bibfield  {journal} {\bibinfo
  {journal} {Phys. Rev. E}\ }\textbf {\bibinfo {volume} {99}},\ \bibinfo
  {pages} {022137} (\bibinfo {year} {2019})}\BibitemShut {NoStop}%
\bibitem [{\citenamefont {Simon}(2009)}]{simon09a}%
  \BibitemOpen
  \bibfield  {author} {\bibinfo {author} {\bibfnamefont {D.}~\bibnamefont
  {Simon}},\ }\bibfield  {title} {\enquote {\bibinfo {title} {Construction of a
  coordinate {B}ethe ansatz for the asymmetric simple exclusion process with
  open boundaries},}\ }\href@noop {} {\bibfield  {journal} {\bibinfo  {journal}
  {J. Stat. Mech. P07017}\ } (\bibinfo {year} {2009})}\BibitemShut {NoStop}%
\bibitem [{\citenamefont {Popkov}\ \emph {et~al.}(2010)\citenamefont {Popkov},
  \citenamefont {Sch{\"u}tz},\ and\ \citenamefont {Simon}}]{popkov10a}%
  \BibitemOpen
  \bibfield  {author} {\bibinfo {author} {\bibfnamefont {V.}~\bibnamefont
  {Popkov}}, \bibinfo {author} {\bibfnamefont {G.~M.}\ \bibnamefont
  {Sch{\"u}tz}}, \ and\ \bibinfo {author} {\bibfnamefont {D.}~\bibnamefont
  {Simon}},\ }\bibfield  {title} {\enquote {\bibinfo {title} {{ASEP} on a ring
  conditioned on enhanced flux},}\ }\href@noop {} {\bibfield  {journal}
  {\bibinfo  {journal} {J. Stat. Mech. P10007}\ } (\bibinfo {year}
  {2010})}\BibitemShut {NoStop}%
\bibitem [{\citenamefont {Jack}\ and\ \citenamefont
  {Sollich}(2010)}]{jack2010}%
  \BibitemOpen
  \bibfield  {author} {\bibinfo {author} {\bibfnamefont {R.~L.}\ \bibnamefont
  {Jack}}\ and\ \bibinfo {author} {\bibfnamefont {P.}~\bibnamefont {Sollich}},\
  }\bibfield  {title} {\enquote {\bibinfo {title} {Large deviations and
  ensembles of trajectories in stochastic models},}\ }\href@noop {} {\bibfield
  {journal} {\bibinfo  {journal} {Prog. Theor. Phys. Supp.}\ }\textbf {\bibinfo
  {volume} {184}},\ \bibinfo {pages} {304} (\bibinfo {year}
  {2010})}\BibitemShut {NoStop}%
\bibitem [{\citenamefont {Chetrite}\ and\ \citenamefont
  {Touchette}(2015)}]{chetrite2015}%
  \BibitemOpen
  \bibfield  {author} {\bibinfo {author} {\bibfnamefont {R.}~\bibnamefont
  {Chetrite}}\ and\ \bibinfo {author} {\bibfnamefont {H.}~\bibnamefont
  {Touchette}},\ }\bibfield  {title} {\enquote {\bibinfo {title}
  {Nonequilibrium {M}arkov processes conditioned on large deviations},}\
  }\bibfield  {booktitle} {\emph {\bibinfo {booktitle} {Ann. Henri
  Poincar{\'e}}},\ }\href@noop {} {\ \textbf {\bibinfo {volume} {16}},\
  \bibinfo {pages} {2005} (\bibinfo {year} {2015})}\BibitemShut {NoStop}%
\bibitem [{\citenamefont {Burda}\ \emph {et~al.}(2009)\citenamefont {Burda},
  \citenamefont {Duda}, \citenamefont {Luck},\ and\ \citenamefont
  {Waclaw}}]{burda2009}%
  \BibitemOpen
  \bibfield  {author} {\bibinfo {author} {\bibfnamefont {Z.}~\bibnamefont
  {Burda}}, \bibinfo {author} {\bibfnamefont {J.}~\bibnamefont {Duda}},
  \bibinfo {author} {\bibfnamefont {J.-M.}\ \bibnamefont {Luck}}, \ and\
  \bibinfo {author} {\bibfnamefont {B.}~\bibnamefont {Waclaw}},\ }\bibfield
  {title} {\enquote {\bibinfo {title} {Localization of the maximal entropy
  random walk},}\ }\href@noop {} {\bibfield  {journal} {\bibinfo  {journal}
  {Phys. Rev. Lett.}\ }\textbf {\bibinfo {volume} {102}},\ \bibinfo {pages}
  {160602} (\bibinfo {year} {2009})}\BibitemShut {NoStop}%
\bibitem [{\citenamefont {Noh}\ and\ \citenamefont {Rieger}(2004)}]{noh2004}%
  \BibitemOpen
  \bibfield  {author} {\bibinfo {author} {\bibfnamefont {J.~D.}\ \bibnamefont
  {Noh}}\ and\ \bibinfo {author} {\bibfnamefont {H.}~\bibnamefont {Rieger}},\
  }\bibfield  {title} {\enquote {\bibinfo {title} {Random walks on complex
  networks},}\ }\href@noop {} {\bibfield  {journal} {\bibinfo  {journal} {Phys.
  Rev. Lett.}\ }\textbf {\bibinfo {volume} {92}},\ \bibinfo {pages} {118701}
  (\bibinfo {year} {2004})}\BibitemShut {NoStop}%
\bibitem [{\citenamefont {Bianconi}(2018)}]{bianconi2018}%
  \BibitemOpen
  \bibfield  {author} {\bibinfo {author} {\bibfnamefont {G.}~\bibnamefont
  {Bianconi}},\ }\bibfield  {title} {\enquote {\bibinfo {title} {Rare events
  and discontinuous percolation transitions},}\ }\href@noop {} {\bibfield
  {journal} {\bibinfo  {journal} {Phys. Rev. E}\ }\textbf {\bibinfo {volume}
  {97}},\ \bibinfo {pages} {022314} (\bibinfo {year} {2018})}\BibitemShut
  {NoStop}%
\bibitem [{\citenamefont {Bianconi}(2019)}]{bianconi2019}%
  \BibitemOpen
  \bibfield  {author} {\bibinfo {author} {\bibfnamefont {G.}~\bibnamefont
  {Bianconi}},\ }\bibfield  {title} {\enquote {\bibinfo {title} {Large
  deviation theory of percolation on multiplex networks},}\ }\href@noop {}
  {\bibfield  {journal} {\bibinfo  {journal} {J. Stat. Mech. P023405}\ }
  (\bibinfo {year} {2019})}\BibitemShut {NoStop}%
\bibitem [{\citenamefont {Hindes}\ and\ \citenamefont
  {Schwartz}(2016)}]{hindes2016}%
  \BibitemOpen
  \bibfield  {author} {\bibinfo {author} {\bibfnamefont {J.}~\bibnamefont
  {Hindes}}\ and\ \bibinfo {author} {\bibfnamefont {I.~B.}\ \bibnamefont
  {Schwartz}},\ }\bibfield  {title} {\enquote {\bibinfo {title} {Epidemic
  extinction and control in heterogeneous networks},}\ }\href@noop {}
  {\bibfield  {journal} {\bibinfo  {journal} {Phys. Rev. Lett.}\ }\textbf
  {\bibinfo {volume} {117}},\ \bibinfo {pages} {028302} (\bibinfo {year}
  {2016})}\BibitemShut {NoStop}%
\bibitem [{\citenamefont {Hindes}\ and\ \citenamefont
  {Schwartz}(2017)}]{hindes2017}%
  \BibitemOpen
  \bibfield  {author} {\bibinfo {author} {\bibfnamefont {J.}~\bibnamefont
  {Hindes}}\ and\ \bibinfo {author} {\bibfnamefont {I.~B.}\ \bibnamefont
  {Schwartz}},\ }\bibfield  {title} {\enquote {\bibinfo {title} {Epidemic
  extinction paths in complex networks},}\ }\href@noop {} {\bibfield  {journal}
  {\bibinfo  {journal} {Phys. Rev. E}\ }\textbf {\bibinfo {volume} {95}},\
  \bibinfo {pages} {052317} (\bibinfo {year} {2017})}\BibitemShut {NoStop}%
\bibitem [{\citenamefont {Hindes}\ and\ \citenamefont
  {Assaf}(2019)}]{hindes2019}%
  \BibitemOpen
  \bibfield  {author} {\bibinfo {author} {\bibfnamefont {J.}~\bibnamefont
  {Hindes}}\ and\ \bibinfo {author} {\bibfnamefont {M.}~\bibnamefont {Assaf}},\
  }\bibfield  {title} {\enquote {\bibinfo {title} {Degree dispersion increases
  the rate of rare events in population networks},}\ }\href@noop {} {\bibfield
  {journal} {\bibinfo  {journal} {Phys. Rev. Lett.}\ }\textbf {\bibinfo
  {volume} {123}},\ \bibinfo {pages} {068301} (\bibinfo {year}
  {2019})}\BibitemShut {NoStop}%
\bibitem [{\citenamefont {Chen}\ \emph {et~al.}(2017)\citenamefont {Chen},
  \citenamefont {Shen}, \citenamefont {Zhang},\ and\ \citenamefont
  {Kurths}}]{chen2017}%
  \BibitemOpen
  \bibfield  {author} {\bibinfo {author} {\bibfnamefont {H.}~\bibnamefont
  {Chen}}, \bibinfo {author} {\bibfnamefont {C.}~\bibnamefont {Shen}}, \bibinfo
  {author} {\bibfnamefont {H.}~\bibnamefont {Zhang}}, \ and\ \bibinfo {author}
  {\bibfnamefont {J{\"u}rgen}\ \bibnamefont {Kurths}},\ }\bibfield  {title}
  {\enquote {\bibinfo {title} {Large deviation induced phase switch in an
  inertial majority-vote model},}\ }\href@noop {} {\bibfield  {journal}
  {\bibinfo  {journal} {Chaos}\ }\textbf {\bibinfo {volume} {27}},\ \bibinfo
  {pages} {081102} (\bibinfo {year} {2017})}\BibitemShut {NoStop}%
\bibitem [{\citenamefont {Chen}\ \emph {et~al.}(2019)\citenamefont {Chen},
  \citenamefont {Huang}, \citenamefont {Li},\ and\ \citenamefont
  {Zhang}}]{chen2019}%
  \BibitemOpen
  \bibfield  {author} {\bibinfo {author} {\bibfnamefont {H.}~\bibnamefont
  {Chen}}, \bibinfo {author} {\bibfnamefont {F.}~\bibnamefont {Huang}},
  \bibinfo {author} {\bibfnamefont {G.}~\bibnamefont {Li}}, \ and\ \bibinfo
  {author} {\bibfnamefont {H.}~\bibnamefont {Zhang}},\ }\bibfield  {title}
  {\enquote {\bibinfo {title} {Large deviation and anomalous fluctuations
  scaling in degree assortativity on configuration networks},}\ }\href@noop {}
  {\bibfield  {journal} {\bibinfo  {journal} {arXiv:1907.13330}\ } (\bibinfo
  {year} {2019})}\BibitemShut {NoStop}%
\bibitem [{\citenamefont {den Hollander}\ \emph {et~al.}(2018)\citenamefont
  {den Hollander}, \citenamefont {Mandjes}, \citenamefont {Roccaverde},\ and\
  \citenamefont {Starreveld}}]{denhollander2018}%
  \BibitemOpen
  \bibfield  {author} {\bibinfo {author} {\bibfnamefont {F.}~\bibnamefont {den
  Hollander}}, \bibinfo {author} {\bibfnamefont {M.}~\bibnamefont {Mandjes}},
  \bibinfo {author} {\bibfnamefont {A.}~\bibnamefont {Roccaverde}}, \ and\
  \bibinfo {author} {\bibfnamefont {N.J.}\ \bibnamefont {Starreveld}},\
  }\bibfield  {title} {\enquote {\bibinfo {title} {Ensemble equivalence for
  dense graphs},}\ }\href@noop {} {\bibfield  {journal} {\bibinfo  {journal}
  {Electron. J. Probab.}\ }\textbf {\bibinfo {volume} {23}},\ \bibinfo {pages}
  {1} (\bibinfo {year} {2018})}\BibitemShut {NoStop}%
\bibitem [{\citenamefont {Giardin{\`a}}\ \emph {et~al.}(2021)\citenamefont
  {Giardin{\`a}}, \citenamefont {Giberti},\ and\ \citenamefont
  {Magnanini}}]{giardina2020}%
  \BibitemOpen
  \bibfield  {author} {\bibinfo {author} {\bibfnamefont {C.}~\bibnamefont
  {Giardin{\`a}}}, \bibinfo {author} {\bibfnamefont {C.}~\bibnamefont
  {Giberti}}, \ and\ \bibinfo {author} {\bibfnamefont {E.}~\bibnamefont
  {Magnanini}},\ }\bibfield  {title} {\enquote {\bibinfo {title} {Approximating
  the cumulant generating function of triangles in the {E}rd\" os-{R}\'enyi
  random graph},}\ }\href@noop {} {\bibfield  {journal} {\bibinfo  {journal}
  {J. Stat. Phys.}\ }\textbf {\bibinfo {volume} {182}},\ \bibinfo {pages} {23}
  (\bibinfo {year} {2021})}\BibitemShut {NoStop}%
\bibitem [{\citenamefont {Metz}\ and\ \citenamefont
  {Castillo}(2019)}]{metz2019}%
  \BibitemOpen
  \bibfield  {author} {\bibinfo {author} {\bibfnamefont {F.~L.}\ \bibnamefont
  {Metz}}\ and\ \bibinfo {author} {\bibfnamefont {I.~P.}\ \bibnamefont
  {Castillo}},\ }\bibfield  {title} {\enquote {\bibinfo {title} {Condensation
  of degrees emerging through a first-order phase transition in classical
  random graphs},}\ }\href@noop {} {\bibfield  {journal} {\bibinfo  {journal}
  {Phys. Rev. E}\ }\textbf {\bibinfo {volume} {100}},\ \bibinfo {pages}
  {012305} (\bibinfo {year} {2019})}\BibitemShut {NoStop}%
\bibitem [{\citenamefont {Masuda}\ \emph {et~al.}(2017)\citenamefont {Masuda},
  \citenamefont {Porter},\ and\ \citenamefont {Lambiotte}}]{masuda2017}%
  \BibitemOpen
  \bibfield  {author} {\bibinfo {author} {\bibfnamefont {N.}~\bibnamefont
  {Masuda}}, \bibinfo {author} {\bibfnamefont {M.~A.}\ \bibnamefont {Porter}},
  \ and\ \bibinfo {author} {\bibfnamefont {R.}~\bibnamefont {Lambiotte}},\
  }\bibfield  {title} {\enquote {\bibinfo {title} {Random walks and diffusion
  on networks},}\ }\href@noop {} {\bibfield  {journal} {\bibinfo  {journal}
  {Phys. Rep.}\ }\textbf {\bibinfo {volume} {716}},\ \bibinfo {pages} {1}
  (\bibinfo {year} {2017})}\BibitemShut {NoStop}%
\bibitem [{\citenamefont {Riascos}\ and\ \citenamefont
  {Mateos}(2019)}]{riascos2019}%
  \BibitemOpen
  \bibfield  {author} {\bibinfo {author} {\bibfnamefont {A.~P.}\ \bibnamefont
  {Riascos}}\ and\ \bibinfo {author} {\bibfnamefont {J.~L.}\ \bibnamefont
  {Mateos}},\ }\bibfield  {title} {\enquote {\bibinfo {title} {Random walks on
  weighted networks: {E}xploring local and non-local navigation strategies},}\
  }\href@noop {} {\bibfield  {journal} {\bibinfo  {journal} {arXiv:1901.05609}\
  } (\bibinfo {year} {2019})}\BibitemShut {NoStop}%
\bibitem [{\citenamefont {Ahuja}\ \emph {et~al.}(1993)\citenamefont {Ahuja},
  \citenamefont {Magnanti},\ and\ \citenamefont {Orlin}}]{ahuja1993}%
  \BibitemOpen
  \bibfield  {author} {\bibinfo {author} {\bibfnamefont {R.K.}\ \bibnamefont
  {Ahuja}}, \bibinfo {author} {\bibfnamefont {T.~L.}\ \bibnamefont {Magnanti}},
  \ and\ \bibinfo {author} {\bibfnamefont {J.~B.}\ \bibnamefont {Orlin}},\
  }\href@noop {} {\emph {\bibinfo {title} {Network {F}lows: {T}heory,
  {A}lgorithms and {A}pplications}}}\ (\bibinfo  {publisher} {Prentice Hall},\
  \bibinfo {address} {Upper Saddle River, NJ},\ \bibinfo {year}
  {1993})\BibitemShut {NoStop}%
\bibitem [{\citenamefont {Budini}\ \emph {et~al.}(2014)\citenamefont {Budini},
  \citenamefont {Turner},\ and\ \citenamefont {Garrahan}}]{budini2014}%
  \BibitemOpen
  \bibfield  {author} {\bibinfo {author} {\bibfnamefont {A.~A.}\ \bibnamefont
  {Budini}}, \bibinfo {author} {\bibfnamefont {R.~M.}\ \bibnamefont {Turner}},
  \ and\ \bibinfo {author} {\bibfnamefont {J.~P.}\ \bibnamefont {Garrahan}},\
  }\bibfield  {title} {\enquote {\bibinfo {title} {Fluctuating observation time
  ensembles in the thermodynamics of trajectories},}\ }\href@noop {} {\bibfield
   {journal} {\bibinfo  {journal} {J. Stat. Mech. P03012}\ } (\bibinfo {year}
  {2014})}\BibitemShut {NoStop}%
\bibitem [{\citenamefont {Garrahan}(2017)}]{garrahan2017}%
  \BibitemOpen
  \bibfield  {author} {\bibinfo {author} {\bibfnamefont {J.~P.}\ \bibnamefont
  {Garrahan}},\ }\bibfield  {title} {\enquote {\bibinfo {title} {Simple bounds
  on fluctuations and uncertainty relations for first-passage times of counting
  observables},}\ }\href@noop {} {\bibfield  {journal} {\bibinfo  {journal}
  {Phys. Rev. E}\ }\textbf {\bibinfo {volume} {95}},\ \bibinfo {pages} {032134}
  (\bibinfo {year} {2017})}\BibitemShut {NoStop}%
\bibitem [{\citenamefont {Carollo}\ \emph {et~al.}(2018)\citenamefont
  {Carollo}, \citenamefont {Garrahan}, \citenamefont {Lesanovsky},\ and\
  \citenamefont {P{\'e}rez-Espigares}}]{Carollo2018}%
  \BibitemOpen
  \bibfield  {author} {\bibinfo {author} {\bibfnamefont {F.}~\bibnamefont
  {Carollo}}, \bibinfo {author} {\bibfnamefont {J.~P.}\ \bibnamefont
  {Garrahan}}, \bibinfo {author} {\bibfnamefont {I.}~\bibnamefont
  {Lesanovsky}}, \ and\ \bibinfo {author} {\bibfnamefont {C.}~\bibnamefont
  {P{\'e}rez-Espigares}},\ }\bibfield  {title} {\enquote {\bibinfo {title}
  {Making rare events typical in {M}arkovian open quantum systems},}\
  }\href@noop {} {\bibfield  {journal} {\bibinfo  {journal} {Phys. Rev. A}\
  }\textbf {\bibinfo {volume} {98}},\ \bibinfo {pages} {010103(R)} (\bibinfo
  {year} {2018})}\BibitemShut {NoStop}%
\bibitem [{\citenamefont {Angeletti}\ and\ \citenamefont
  {Touchette}(2016)}]{angeletti16a}%
  \BibitemOpen
  \bibfield  {author} {\bibinfo {author} {\bibfnamefont {F.}~\bibnamefont
  {Angeletti}}\ and\ \bibinfo {author} {\bibfnamefont {H.}~\bibnamefont
  {Touchette}},\ }\bibfield  {title} {\enquote {\bibinfo {title} {Diffusions
  conditioned on occupation measures},}\ }\href@noop {} {\bibfield  {journal}
  {\bibinfo  {journal} {J. Math. Phys.}\ }\textbf {\bibinfo {volume} {57}},\
  \bibinfo {pages} {023303} (\bibinfo {year} {2016})}\BibitemShut {NoStop}%
\bibitem [{\citenamefont {Villavicencio-Sanchez}\ and\ \citenamefont
  {Harris}(2016)}]{villavicencio16a}%
  \BibitemOpen
  \bibfield  {author} {\bibinfo {author} {\bibfnamefont {R.}~\bibnamefont
  {Villavicencio-Sanchez}}\ and\ \bibinfo {author} {\bibfnamefont {R.~J.}\
  \bibnamefont {Harris}},\ }\bibfield  {title} {\enquote {\bibinfo {title}
  {Local structure of current fluctuations in diffusive systems beyond one
  dimension},}\ }\href@noop {} {\bibfield  {journal} {\bibinfo  {journal}
  {Phys. Rev. E}\ }\textbf {\bibinfo {volume} {93}},\ \bibinfo {pages} {032134}
  (\bibinfo {year} {2016})}\BibitemShut {NoStop}%
\bibitem [{\citenamefont {Bodineau}\ \emph {et~al.}(2008)\citenamefont
  {Bodineau}, \citenamefont {Derrida},\ and\ \citenamefont
  {Lebowitz}}]{bodineau2008}%
  \BibitemOpen
  \bibfield  {author} {\bibinfo {author} {\bibfnamefont {T.}~\bibnamefont
  {Bodineau}}, \bibinfo {author} {\bibfnamefont {B.}~\bibnamefont {Derrida}}, \
  and\ \bibinfo {author} {\bibfnamefont {J.~L.}\ \bibnamefont {Lebowitz}},\
  }\bibfield  {title} {\enquote {\bibinfo {title} {Vortices in the
  two-dimensional simple exclusion process},}\ }\href@noop {} {\bibfield
  {journal} {\bibinfo  {journal} {J. Stat. Phys.}\ }\textbf {\bibinfo {volume}
  {131}},\ \bibinfo {pages} {821} (\bibinfo {year} {2008})}\BibitemShut
  {NoStop}%
\bibitem [{\citenamefont {Villavicencio-Sanchez}\ \emph
  {et~al.}(2012)\citenamefont {Villavicencio-Sanchez}, \citenamefont {Harris},\
  and\ \citenamefont {Touchette}}]{villavicencio12a}%
  \BibitemOpen
  \bibfield  {author} {\bibinfo {author} {\bibfnamefont {R.}~\bibnamefont
  {Villavicencio-Sanchez}}, \bibinfo {author} {\bibfnamefont {R.~J.}\
  \bibnamefont {Harris}}, \ and\ \bibinfo {author} {\bibfnamefont
  {H.}~\bibnamefont {Touchette}},\ }\bibfield  {title} {\enquote {\bibinfo
  {title} {Current loops and fluctuations in the zero-range process on a
  diamond lattice},}\ }\href@noop {} {\bibfield  {journal} {\bibinfo  {journal}
  {J. Stat. Mech. P07007}\ } (\bibinfo {year} {2012})}\BibitemShut {NoStop}%
\bibitem [{\citenamefont {Giardin\`a}\ \emph {et~al.}(2006)\citenamefont
  {Giardin\`a}, \citenamefont {Kurchan},\ and\ \citenamefont
  {Peliti}}]{giardina06a}%
  \BibitemOpen
  \bibfield  {author} {\bibinfo {author} {\bibfnamefont {C.}~\bibnamefont
  {Giardin\`a}}, \bibinfo {author} {\bibfnamefont {J.}~\bibnamefont {Kurchan}},
  \ and\ \bibinfo {author} {\bibfnamefont {L.}~\bibnamefont {Peliti}},\
  }\bibfield  {title} {\enquote {\bibinfo {title} {Direct evaluation of
  large-deviation functions},}\ }\href@noop {} {\bibfield  {journal} {\bibinfo
  {journal} {Phys. Rev. Lett.}\ }\textbf {\bibinfo {volume} {96}},\ \bibinfo
  {pages} {120603} (\bibinfo {year} {2006})}\BibitemShut {NoStop}%
\bibitem [{\citenamefont {Giardin\`a}\ \emph {et~al.}(2011)\citenamefont
  {Giardin\`a}, \citenamefont {Kurchan}, \citenamefont {Lecomte},\ and\
  \citenamefont {Tailleur}}]{giardina11a}%
  \BibitemOpen
  \bibfield  {author} {\bibinfo {author} {\bibfnamefont {C.}~\bibnamefont
  {Giardin\`a}}, \bibinfo {author} {\bibfnamefont {J.}~\bibnamefont {Kurchan}},
  \bibinfo {author} {\bibfnamefont {V.}~\bibnamefont {Lecomte}}, \ and\
  \bibinfo {author} {\bibfnamefont {J.}~\bibnamefont {Tailleur}},\ }\bibfield
  {title} {\enquote {\bibinfo {title} {Simulating rare events in dynamical
  processes},}\ }\href@noop {} {\bibfield  {journal} {\bibinfo  {journal} {J.
  Stat. Phys.}\ }\textbf {\bibinfo {volume} {145}},\ \bibinfo {pages}
  {787--811} (\bibinfo {year} {2011})}\BibitemShut {NoStop}%
\bibitem [{\citenamefont {Carollo}\ and\ \citenamefont
  {P\'erez-Espigares}(2020)}]{carollo20a}%
  \BibitemOpen
  \bibfield  {author} {\bibinfo {author} {\bibfnamefont {F.}~\bibnamefont
  {Carollo}}\ and\ \bibinfo {author} {\bibfnamefont {C.}~\bibnamefont
  {P\'erez-Espigares}},\ }\bibfield  {title} {\enquote {\bibinfo {title}
  {Entanglement statistics in {M}arkovian open quantum systems: {A} matter of
  mutation and selection},}\ }\href@noop {} {\bibfield  {journal} {\bibinfo
  {journal} {Phys. Rev. E}\ }\textbf {\bibinfo {volume} {102}},\ \bibinfo
  {pages} {030104(R)} (\bibinfo {year} {2020})}\BibitemShut {NoStop}%
\bibitem [{\citenamefont {Ferr{\'e}}\ and\ \citenamefont
  {Touchette}(2018)}]{ferre18a}%
  \BibitemOpen
  \bibfield  {author} {\bibinfo {author} {\bibfnamefont {G.}~\bibnamefont
  {Ferr{\'e}}}\ and\ \bibinfo {author} {\bibfnamefont {H.}~\bibnamefont
  {Touchette}},\ }\bibfield  {title} {\enquote {\bibinfo {title} {Adaptive
  sampling of large deviations},}\ }\href@noop {} {\bibfield  {journal}
  {\bibinfo  {journal} {J. Stat. Phys.}\ }\textbf {\bibinfo {volume} {172}},\
  \bibinfo {pages} {1525--1544} (\bibinfo {year} {2018})}\BibitemShut {NoStop}%
\bibitem [{\citenamefont {Rose}\ \emph {et~al.}(2021)\citenamefont {Rose},
  \citenamefont {Mair},\ and\ \citenamefont {Garrahan}}]{rose20a}%
  \BibitemOpen
  \bibfield  {author} {\bibinfo {author} {\bibfnamefont {D.~C.}\ \bibnamefont
  {Rose}}, \bibinfo {author} {\bibfnamefont {J.~F.}\ \bibnamefont {Mair}}, \
  and\ \bibinfo {author} {\bibfnamefont {J.~P.}\ \bibnamefont {Garrahan}},\
  }\bibfield  {title} {\enquote {\bibinfo {title} {A reinforcement learning
  approach to rare trajectory sampling},}\ }\href@noop {} {\bibfield  {journal}
  {\bibinfo  {journal} {New J. Phys.}\ }\textbf {\bibinfo {volume} {23}},\
  \bibinfo {pages} {013013} (\bibinfo {year} {2021})}\BibitemShut {NoStop}%
\bibitem [{\citenamefont {Whitelam}\ and\ \citenamefont
  {Tamblyn}(2020)}]{whitelam20a}%
  \BibitemOpen
  \bibfield  {author} {\bibinfo {author} {\bibfnamefont {S.}~\bibnamefont
  {Whitelam}}\ and\ \bibinfo {author} {\bibfnamefont {I.}~\bibnamefont
  {Tamblyn}},\ }\bibfield  {title} {\enquote {\bibinfo {title} {Learning to
  grow: {C}ontrol of material self-assembly using evolutionary reinforcement
  learning},}\ }\href@noop {} {\bibfield  {journal} {\bibinfo  {journal} {Phys.
  Rev. E}\ }\textbf {\bibinfo {volume} {101}},\ \bibinfo {pages} {052604}
  (\bibinfo {year} {2020})}\BibitemShut {NoStop}%
\bibitem [{\citenamefont {Barr}\ \emph {et~al.}(2020)\citenamefont {Barr},
  \citenamefont {Gispen},\ and\ \citenamefont {Lamacraft}}]{barr20a}%
  \BibitemOpen
  \bibfield  {author} {\bibinfo {author} {\bibfnamefont {A.}~\bibnamefont
  {Barr}}, \bibinfo {author} {\bibfnamefont {W.}~\bibnamefont {Gispen}}, \ and\
  \bibinfo {author} {\bibfnamefont {A.}~\bibnamefont {Lamacraft}},\ }\bibfield
  {title} {\enquote {\bibinfo {title} {Quantum ground states from reinforcement
  learning},}\ }in\ \href@noop {} {\emph {\bibinfo {booktitle} {Proceedings of
  The First Mathematical and Scientific Machine Learning Conference}}},\
  \bibinfo {series} {Proceedings of Machine Learning Research}, Vol.\ \bibinfo
  {volume} {107},\ \bibinfo {editor} {edited by\ \bibinfo {editor}
  {\bibfnamefont {Jianfeng}\ \bibnamefont {Lu}}\ and\ \bibinfo {editor}
  {\bibfnamefont {Rachel}\ \bibnamefont {Ward}}}\ (\bibinfo  {publisher}
  {PMLR},\ \bibinfo {address} {Princeton University, Princeton, NJ, USA},\
  \bibinfo {year} {2020})\ pp.\ \bibinfo {pages} {635--653}\BibitemShut
  {NoStop}%
\bibitem [{\citenamefont {{Iannone}}\ \emph {et~al.}(2019)\citenamefont
  {{Iannone}}, \citenamefont {{Ambrosino}}, \citenamefont {{Bracco}},
  \citenamefont {{De Rosa}}, \citenamefont {{Funel}}, \citenamefont
  {{Guarnieri}}, \citenamefont {{Migliori}}, \citenamefont {{Palombi}},
  \citenamefont {{Ponti}}, \citenamefont {{Santomauro}},\ and\ \citenamefont
  {{Procacci}}}]{iannone2019}%
  \BibitemOpen
  \bibfield  {author} {\bibinfo {author} {\bibfnamefont {F.}~\bibnamefont
  {{Iannone}}}, \bibinfo {author} {\bibfnamefont {F.}~\bibnamefont
  {{Ambrosino}}}, \bibinfo {author} {\bibfnamefont {G.}~\bibnamefont
  {{Bracco}}}, \bibinfo {author} {\bibfnamefont {M.}~\bibnamefont {{De Rosa}}},
  \bibinfo {author} {\bibfnamefont {A.}~\bibnamefont {{Funel}}}, \bibinfo
  {author} {\bibfnamefont {G.}~\bibnamefont {{Guarnieri}}}, \bibinfo {author}
  {\bibfnamefont {S.}~\bibnamefont {{Migliori}}}, \bibinfo {author}
  {\bibfnamefont {F.}~\bibnamefont {{Palombi}}}, \bibinfo {author}
  {\bibfnamefont {G.}~\bibnamefont {{Ponti}}}, \bibinfo {author} {\bibfnamefont
  {G.}~\bibnamefont {{Santomauro}}}, \ and\ \bibinfo {author} {\bibfnamefont
  {P.}~\bibnamefont {{Procacci}}},\ }\bibfield  {title} {\enquote {\bibinfo
  {title} {{CRESCO ENEA HPC} clusters: a working example of a multifabric
  {GPFS} {S}pectrum {S}cale layout},}\ }in\ \href@noop {} {\emph {\bibinfo
  {booktitle} {2019 {I}nternational {C}onference on {H}igh {P}erformance
  {C}omputing {S}imulation ({HPCS})}}}\ (\bibinfo {year} {2019})\ p.\ \bibinfo
  {pages} {1051}\BibitemShut {NoStop}%
\bibitem [{\citenamefont {Grimmett}\ and\ \citenamefont
  {Stirzaker}(1992)}]{grimmett1992}%
  \BibitemOpen
  \bibfield  {author} {\bibinfo {author} {\bibfnamefont {G.~R.}\ \bibnamefont
  {Grimmett}}\ and\ \bibinfo {author} {\bibfnamefont {D.~R.}\ \bibnamefont
  {Stirzaker}},\ }\href@noop {} {\emph {\bibinfo {title} {Probability and
  {R}andom Processes}}}\ (\bibinfo  {publisher} {Clarendon Press},\ \bibinfo
  {address} {Oxford},\ \bibinfo {year} {1992})\BibitemShut {NoStop}%
\bibitem [{\citenamefont {G{\'o}mez-Gardenes}\ and\ \citenamefont
  {Latora}(2008)}]{gomezgardenes2008}%
  \BibitemOpen
  \bibfield  {author} {\bibinfo {author} {\bibfnamefont {J.}~\bibnamefont
  {G{\'o}mez-Gardenes}}\ and\ \bibinfo {author} {\bibfnamefont
  {V.}~\bibnamefont {Latora}},\ }\bibfield  {title} {\enquote {\bibinfo {title}
  {Entropy rate of diffusion processes on complex networks},}\ }\href@noop {}
  {\bibfield  {journal} {\bibinfo  {journal} {Phys. Rev. E}\ }\textbf {\bibinfo
  {volume} {78}},\ \bibinfo {pages} {065102(R)} (\bibinfo {year}
  {2008})}\BibitemShut {NoStop}%
\bibitem [{\citenamefont {Kishore}\ \emph {et~al.}(2012)\citenamefont
  {Kishore}, \citenamefont {Santhanam},\ and\ \citenamefont
  {Amritkar}}]{kishore2012}%
  \BibitemOpen
  \bibfield  {author} {\bibinfo {author} {\bibfnamefont {V.}~\bibnamefont
  {Kishore}}, \bibinfo {author} {\bibfnamefont {M.~S.}\ \bibnamefont
  {Santhanam}}, \ and\ \bibinfo {author} {\bibfnamefont {R.~E.}\ \bibnamefont
  {Amritkar}},\ }\bibfield  {title} {\enquote {\bibinfo {title} {Extreme events
  and event size fluctuations in biased random walks on networks},}\
  }\href@noop {} {\bibfield  {journal} {\bibinfo  {journal} {Phys. Rev. E}\
  }\textbf {\bibinfo {volume} {85}},\ \bibinfo {pages} {056120} (\bibinfo
  {year} {2012})}\BibitemShut {NoStop}%
\bibitem [{\citenamefont {Monthus}(2011)}]{monthus2011}%
  \BibitemOpen
  \bibfield  {author} {\bibinfo {author} {\bibfnamefont {C.}~\bibnamefont
  {Monthus}},\ }\bibfield  {title} {\enquote {\bibinfo {title} {Non-equilibrium
  steady states: maximization of the {S}hannon entropy associated with the
  distribution of dynamical trajectories in the presence of constraints},}\
  }\href@noop {} {\bibfield  {journal} {\bibinfo  {journal} {J. Stat. Mech.
  P03008}\ } (\bibinfo {year} {2011})}\BibitemShut {NoStop}%
\end{thebibliography}%

%\newpage

\appendix

\section{Random walks on networks}

\renewcommand{\thesubsection}{A\arabic{subsection}}

We consider two different types of discrete-time random walks on networks: the standard random walk \cite{noh2004} and the maximal entropy random walk \cite{burda2009}. Below we will explain why these choices are particularly well adapted to the problems at hand ---others could conceivably be better addressed by other random-walk local rules (see e.g. \cite{riascos2019} and references therein). The networks are of finite size $N$, directed and strongly connected, i.e.\! a walker can reach any of the $N$ nodes from any other node along directed links \cite{newman2010}. Connected undirected networks are also implicitly considered as a particular case.

\subsection{Standard random walk}

The standard random walk (SRW) ---variously known in the literature as generic, unbiased, normal, uniform random walk, or simply {\it the} random walk--- has been extensively studied in the last two decades  \cite{noh2004,masuda2017,riascos2019}. A random walker moves every time step to one of the neighboring nodes that can be reached from its present location, the specific destination being chosen (uniformly) at random among the different possibilities. Given a network with directed adjacency matrix ${\bf A}$, where $A_{ji} = 1$ if there is a link pointing from $i$ to $j$ and is $0$ otherwise, the transition matrix of the SRW, ${\bf \Pi}^{\text{SRW}}$, thus assigns the same probability to each of the directed links emanating from a given node
\begin{equation}
\Pi_{ji}^{\text{SRW}} = \frac{A_{ji}}{k^\text{out}_i}\,.
\label{swr}
\end{equation}
The normalization by the out-degree of node $i$, defined as the number of neighbors joined by outgoing links $k^\text{out}_i = \sum_j A_{ji}$, ensures the conservation of probability, $\sum_j \Pi^{\text{SRW}}_{ji} = 1$. If the probability of occupying node $i$ at time step $n$ is denoted by $p_i(n)$, then this $N$-state process evolves according to $p_j(n+1) =  \sum_i \Pi^{\text{SRW}}_{ji}\, p_i(n)$. Due to general properties of finite Markov chains \cite{grimmett1992}, the systems evolves asymptotically towards a single stationary state, ${\bm \pi}$, $\lim_{n\to\infty} ({\bf \Pi}^{\text{SRW}})^n {\bm p}(0) = {\bm \pi}$ ---where  $ {\bm p}(0)$ is an arbitrary initial state---, which is the right eigenvector of the transition matrix associated with eigenvalue $1$, ${\bf \Pi}^{\text{SRW}} {\bm\pi}={\bm \pi} $. While the SRW on undirected networks is reversible and the probability of occupying a node in the stationary state is proportional to its degree, for directed networks detailed balance is generally not satisfied and the stationary state in general can only be found approximately \cite{masuda2017}. When considering particle currents or other types of flows, the SRW appears to be the most natural choice, as illustrated in Section IV.

\subsection{Maximal entropy random walk}

In a regular network, where all nodes have the same out-degree, every walk comprising a given number of jumps  governed by Eq.~(\ref{swr})  occurs with the same probability. In a more general setting, however, sequences of nodes visited by the walker of the same length, joining a given source node to a given target node, are not equiprobable. In fact, the SRW trajectory $\omega_{\tau}= (i_\tau \leftarrow \cdots \leftarrow i_2 \leftarrow i_1 \leftarrow i_0)$, occurs with probability $P(\omega_{\tau})=p_{i_0}(0)/(k^\text{out}_{i_\tau}\cdots k^\text{out}_{i_1} k^\text{out}_{i_0})$, which in general depends on the out-degree of the intermediate nodes. 

In order to explore generalized optimal paths in networks (see Section III), it is necessary to start from a Markov chain that assigns the same probability to all walks of the same number of steps joining a given pair of nodes. The goal is that the contributions of specific walks to the solution only depend on the statistics of the observables under study and not on the out-degree of the visited nodes. For this reason we focus on the maximal entropy random walk (MERW) \cite{burda2009}, whose transition matrix is 
\begin{equation}
\Pi_{ji}^\text{MERW} = \frac{A_{ji}}{\lambda}\frac{v_j}{v_i}.
\label{mewr}
\end{equation}
where $\lambda$ is the largest eigenvalue of the directed adjacency matrix, and ${\bm v}$ is the normalized eigenvector associated to it, ${\bm A}{\bm v} = \lambda {\bm v}$. In fact,  ${\bm v}$ is the eigenvector centrality \cite{newman2010}, so the MERW can be considered as a biased random walk \cite{gomezgardenes2008, kishore2012, riascos2019} based on this node-centrality measure. This type of random walk assigns the same probability to each walk of $\tau$ steps between the source $i_0$ and the target $i_\tau$, namely  $\displaystyle \frac{p_{i_0}(0)}{\lambda^\tau}\frac{v_{i_\tau}}{v_{i_0}}$. 

Moreover, it can be shown that the MERW produces Shannon entropy at a rate  $\ln \lambda$, which is the highest possible entropy-production rate for a discrete-time random walk \cite{burda2009}. While the SRW maximizes the entropy locally (i.e.\! in a single jump), the MERW optimizes the entropy along a trajectory, and in fact it has been studied in a quite general framework of dynamic entropy maximization with constraints \cite{monthus2011}.  For undirected networks the stationary-state probability of a given node is the square of its eigenvector centrality, while for directed networks one encounters difficulties similar to those that arise in the characterization of the stationary state of the SRW (see above).  At the end of Appendix C we briefly comment on a connection between the MERW and the SRW that has been recently unveiled.

\section{Ensembles of trajectories}

\renewcommand{\thesubsection}{B\arabic{subsection}}

Time-integrated observables of random walks on networks are studied by means of a thermodynamic formalism of trajectories or histories \cite{lecomte2007}. Specifically, for a trajectory $\omega_{\tau}= (i_\tau \leftarrow \cdots \leftarrow i_2 \leftarrow i_1 \leftarrow i_0)$, involving $\tau \in \mathbb{N}$ jumps, we consider time-extensive observables of the form $\hat{O}(\omega_{\tau}) = \sum_{n=1}^\tau \hat{o}(i_n \leftarrow i_{n-1})$, where $\hat{o}(i_n \leftarrow i_{n-1})$ is the increment of the observable at each time step, whose value depends on the nodes $i_{n-1}$ and $i_n$, which are joined by a link.  While the probability associated to a given trajectory  is just $P(\omega_{\tau}) =\Pi_{i_{\tau} i_{\tau-1}} \cdots \Pi_{i_2 i_1}\Pi_{i_1 i_0}p_{i_0}(0)$, where the details of the transition matrix ${\bf \Pi}$ depend on the choice of random walk, the probability distribution of the observable is $P_{\tau}(O) = \sum_{\omega_{\tau}}\delta(O - \hat{O}(\omega_{\tau})) P(\omega_{\tau})$. This distribution corresponds to an ensemble of trajectories (walks) with fixed observable $O$ and fixed time $\tau$. If the random walk is ergodic, it acquires a large-deviation form, which in terms of the time-intensive observable $O/\tau$ is $P_{\tau}(O)\sim e^{-\tau I(O/\tau)}$, where the function $I(O/\tau)$ is called the rate function, and plays the role of a dynamical entropy. 

This ensemble of trajectories is analogous to the micro-canonical ensemble of configurations in equilibrium statistical mechanics, and, also in this context, it is less useful to work with than other ensembles that  yield equivalent results in the $\tau \to\infty$ limit. In the following we will briefly review the thermodynamics of trajectories approach to the study of the large deviations of such time-integrated observables. The main theoretical framework is developed in Ref.~\cite{garrahan2009}, though some of the developments that we use are more recent ---appropriate references are cited below. 

\subsection{$s$-ensemble}

By biasing each trajectory with a tilting parameter $s$ conjugate to the observable $O$, we obtain the $s$-ensemble $P^s_{\tau}(O) = Z_{\tau}^{-1}(s)\,  e^{-s O}P_{\tau}(O)$, where the normalization factor is a dynamical partition function $Z_{\tau}(s) = \sum_O e^{-s O} P_{\tau}(O)$. In this ensemble, $\tau$ is still fixed, but $O$ is not, as it can fluctuate ---instead, the tilting parameter $s$, which plays a role similar to an inverse temperature, is fixed. For large $\tau$, the partition function also acquires a large-deviation form,  $Z_{\tau}(s) \sim e^{\tau \theta(s)}$, where the so-called scaled-cumulant generating function (SCGF) $\theta(s)$ is related to the rate function defined above by a Legendre-Fenchel transform.  The SCGF can be seen as a dynamical free energy, whose derivates yield the cumulants of $O$,
\begin{equation}
\lim_{\tau\to\infty}\frac{\langle\langle O^p \rangle\rangle_s}{\tau} = (-1)^p \frac{d^p  \theta(s)}{d s^p},
\end{equation}
$\langle\langle O^p \rangle\rangle_s$ being the $p$-th order cumulant of $O$ for a certain value of $s$. We focus on the first and second derivatives, corresponding to the average value and the fluctuations. While these derivatives evaluated at $s=0$ yield a statistical characterization of the natural dynamics, we will also consider them for $s\neq 0$, which (as shown below) corresponds to a dynamics that does not conserve probability. In Appendix C we explain how to obtain proper stochastic dynamics that yield the same statistics for $O$ that are found for $s\neq 0$.

It turns out that the partition function can be written as $Z_{\tau}(s) = \sum_{i,j}  [\left({\bm \Pi^s}\right)^\tau]_{ji} p_{i}(0)$, where the transfer matrix  ${\bf \Pi^s}$ is the tilted operator, whose matrix elements are
\begin{equation}
 \Pi_{ji}^s = e^{-s \hat{o}(j \leftarrow i)}\Pi_{ji}
\label{tiltedsapp}
\end{equation}
for a given random-walk transition matrix $\Pi_{ji}$. The largest eigenvalue of this matrix corresponds to $e^{\theta(s)}$ \cite{garrahan2009}. Finding the statistics of $O$ thus becomes an eigenvalue problem of the tilted generator, which is not a stochastic generator, $\sum_j \Pi_{ji}^s \neq 1$ for $s\neq 0$.

\subsection{$ss$-ensemble}

If we consider two time-integrated observables instead of one, let them be denoted as $O_1$ and $O_2$, which can fluctuate in a trajectory of fixed duration $\tau$,  then we are in the $ss$-ensemble,  $P^{s_1 s_2}_{\tau}(O_1,O_2) = Z_{\tau}^{-1}(s_1,s_2)\,  e^{-s_1 O_1 - s_2 O_2}P_{\tau}(O_1,O_2)$. Here $s_1$ and $s_2$ are two tilting parameters conjugate to $O_1$ and $O_2$, respectively. The $ss$-ensemble is the ensemble of choice for the study of constrained flows in networks developed in Section IV. The large-deviation form of the partition function is in this case $Z_{\tau}(s_1,s_2) \sim e^{\tau \theta(s_1,s_2)}$, and it can be obtained again by computing the largest eigenvalue of the tilted operator, which adopts the form
\begin{equation}
 \Pi_{ji}^{s_1s_2} = e^{-s_1 \hat{o}_1(j\leftarrow i) -s_2 \hat{o}_2(j\leftarrow i)} \Pi_{ji}.
\label{tilted}
\end{equation}
From the (first, second, or higher-order) derivatives of  $\theta(s_1,s_2)$ with respect to $s_1$ or $s_2$ only, one obtains the cumulants of $O_1$ and $O_2$. The cross derivatives yield the $O_1 O_2$-correlations: 
\begin{equation}
\displaystyle\lim_{\tau\to\infty}\frac{\langle \Delta O_1 \Delta O_2 \rangle_{s_1 s_2}}{\tau} = \frac{\partial^2  \theta(s_1,s_2)}{\partial s_1 \partial s_2},
\end{equation}
where $\Delta O_1 = O_1 - \langle O_1\rangle_{s_1 s_2}$, and $\Delta O_2$ is defined analogously.

\subsection{$x$-ensemble}

If instead of keeping fixed the duration of a trajectory $\tau$,  we fix the value that the observable $O$ reaches for each trajectory ---while allowing $\tau$ to fluctuate---, we obtain a different statistical ensemble, namely the $x$-ensemble  \cite{budini2014}. We focus the description of this ensemble on the case in which the observable is a local activity that equals one if a given link (or set of links) is traversed, or zero otherwise, as this is the case that we explore in Section III. What follows is a discrete-time version of the discussion to be found in Section IV of Ref.~\cite{garrahan2017}.

The probability that the observable reaches a given value $O$ at time $\tau$ is the probability that the link or links that contribute to $O$ are traversed at times $\tau_1, \tau_2,\ldots, \tau_{O-1}, \tau_O = \tau$, where all but the last one can take any value. In terms of the operator ${\bf \Pi}_O$, which preserves the transition probabilities of ${\bf \Pi}$ only for those transitions (links) contributing to $O$, and  ${\bf \tilde \Pi}={\bf \Pi}-{\bf \Pi}_O$, this probability is given by
\begin{equation}
P_O(\tau)\!=\!\sum_{ji}\!\sum_{0\leq \tau_1 \leq \cdots \leq \tau}\![ {\bf \Pi}_O {\bf \tilde \Pi}^{\tau-\tau_{O-1}-1}\cdots {\bf \Pi}_O{\bf \tilde \Pi}^{\tau_1-1}]_{ji} p_{i}(0).
\label{px}
\end{equation}
The $x$-ensemble probability distribution is $P_O^x(\tau) =Z_{O}(x)^{-1}\, e^{-x \tau} P_O(\tau)$, where the normalization factor is
\begin{widetext}
\begin{eqnarray}
Z_{O}(x)\!&=&\sum_{\tau=0} ^\infty e^{-x \tau} P_O(\tau) = \sum_{ji} \sum_{\tau=0} ^\infty e^{-x \tau} \sum_{0\leq \tau_1 \cdots \leq \tau} [{\bf \Pi}_O {\bf \tilde \Pi}^{\tau-\tau_{O-1}-1}\cdots {\bf \Pi}_O{\bf \tilde \Pi}^{\tau_1-1}]_{ji}\, p_{i}(0) \nonumber\\
&=& \sum_{ji}  \sum_{\Delta \tau_1=0}^\infty\cdots \sum_{\Delta \tau_O=0}^\infty \left[e^{-x}{\bf \Pi}_O (e^{-x}{\bf \tilde \Pi})^{\Delta \tau_O}\cdots e^{-x}{\bf \Pi}_O(e^{-x}{\bf \tilde \Pi})^{\Delta \tau_1}\right]_{ji} p_{i}(0)\nonumber\\
&=& \sum_{ji} \left[\left( e^{-x}{\bf \Pi}_O \sum_{\Delta \tau=0}^\infty (e^{-x}{\bf \tilde \Pi})^{\Delta \tau}\right)^O\right]_{ji} p_{i}(0) = \sum_{ji} [({\bm \Pi}^x)^O]_{ji} p_{i}(0).
\label{zx}
\end{eqnarray}
\end{widetext}
The auxiliary time-increment variables $\Delta \tau_1 = \tau_1-1$, $\Delta \tau_2 = \tau_2-\tau_1-1$,$\ldots$, $\Delta\tau_0= \tau-\tau_{O-1}-1$, whose values are unrestricted due to the sum over $\tau$, are introduced in the second line so as to emphasize that there are just $O$ identical factors of the form $e^{-x}{\bf \Pi}_O \sum_{\Delta \tau=0}^\infty (e^{-x}{\bf \tilde \Pi})^{\Delta \tau}$. Once the sum over $\Delta \tau$ is performed (see the discussion below), the result is a dynamical grand-partition function written in terms of a transfer operator
\begin{equation}
{\bm \Pi^x} = {\bf \Pi}_O\, (e^{x}- {\bf \tilde \Pi} )^{-1}\, .
\label{transferx}
\end{equation}
This is the tilted operator for the $x$-ensemble.

For large $O$, which also corresponds to large $\tau$, the grand-partition function acquires a large-deviation form $Z_O(x) \sim  e^{O \varphi(x)}$, where $e^{\varphi(x)}$ is the largest eigenvalue of ${\bm \Pi^x}$. The cumulants of the fluctuating time $\tau$ can then be obtained from the derivatives of $\varphi(x)$:
\begin{equation}
\lim_{O\to\infty}\frac{\langle\langle \tau^p \rangle\rangle_x}{O} = (-1)^p \frac{d^p  \varphi(x)}{d x^p}.
\end{equation}
 This ensemble is equivalent to the $s$-ensemble in the large $\tau$ (large $O$) limit  ---see Refs.~\cite{budini2014,garrahan2017}, whose discussion for continuous-time dynamics can be easily adapted to discrete time. Nevertheless, for finding optimal paths the $x$-ensemble is more natural, as one is interested in finding the number of steps (i.e.\! the fluctuating time) needed to reach certain target nodes. The  length of the walk between consecutive observable increments, $\tau/O$ for fixed and large number of repetitions $O$, is more appropriate for the task than the $s$-ensemble $O/\tau$, for fixed and large $\tau$, even if one can be easily obtained from the other. A simple mapping relating the large-deviation functions of one and the other ensemble exists ---see Ref.~\cite{budini2014}, as well as the pertinent discussion in Appendix C.

Since the transfer operator in Eq.~(\ref{transferx}) results from the sum of the matrix series $ \sum_{\Delta \tau=0}^\infty (e^{-x}{\bf \tilde \Pi})^{\Delta \tau}$, and this sum only converges for $x > \log(\lambda_\textrm{max})$,  where $\lambda_\textrm{max}$ is the largest eigenvalue of ${\bf \tilde \Pi}$, only in that range is the SCGF well defined. As we move along the $x$-axis towards smaller values and we get sufficiently close to $\log(\lambda_\textrm{max})$ we see a divergence in the SCGF, and consequently in (minus) its first derivative $\langle \tau\rangle/O$. In Section III the physical meaning of such divergences is clarified. A divergence is clearly observed in Fig.~\ref{fig1} (d) ---and also in Fig.~\ref{fig2} (a) and (b), in the case of the $sx$-ensemble, see below---, and explained in the text that discusses it. In the case of Fig.~\ref{fig1} (c), there is no such divergence, as the largest eigenvalue of ${\bf \tilde \Pi}$ for the ring with a shortcut is $\lambda_\textrm{max}=0$. Physically, it is obvious that no divergence is possible there, as the red node is always reached.

\subsection{$sx$-ensemble}

We next consider an ensemble involving two observables $O$ and $K$ in a trajectory of duration $\tau$, like in the $ss$-ensemble. In this case, however, $O$ is fixed (i.e.\! the trajectory ends when $O$ reaches a certain value) and $K$ is a fluctuating observable that is a local activity. Moreover, the duration $\tau$ is also fluctuating. This is the $sx$-ensemble, and to the best of our knowledge, it has not been studied previously. The tilted probability in this ensemble has two tilting parameters $s$ and $x$
\begin{equation}
P^{s x}_O(K,\tau) = Z_O^{-1}(s,x)\,  e^{- s K -x \tau}P_O(K,\tau)
\end{equation}
where $P_O(K,\tau)$ is the probability that an unbiased trajectory has a given duration $\tau$ and the fluctuating observable reaches a certain value $K$ by the time the other observable reaches its fixed value $O$. The corresponding grand-partition function is
\begin{equation}
Z_O(s,x) = \sum_{K} \sum_\tau  e^{- s K -x \tau}P_O(K,\tau) \sim e^{O \varphi(s,x)},
\end{equation}
where the right-hand side takes the expected large-deviation form with  SCGF $\varphi(s,x)$.

We again obtain the SCGF $\varphi(s,x)$ from the largest eigenvalue of a transfer operator. To do so, this time we need to split the original transition matrix into three parts
\begin{equation}
{\bf \Pi}  = {\bf \Pi}_O + {\bf \Pi}_K +   {\bf \hat \Pi}
\end{equation}
where the three terms of the right-hand side include all transitions that result in an update of observable $O$, all transitions that result in an update of $K$, and the rest of the transitions, respectively. We obtain
\begin{widetext}
\begin{eqnarray}
Z_O(s,x) &=&  \sum_{K} \sum_{\tau=0}^{\infty} e^{- s K}  e^{-x \tau} \sum_{ji}\! \,\sum_{0\leq \tau_1 \leq \cdots \leq \tau}\!   [{\bf \Pi}_O {\bf \hat \Pi}^{\tau-\tau_{O+K-1}-1} {\bf \Pi}_K\cdots  {\bf \Pi}_O {\bf \hat \Pi}^{\tau_2-\tau_1-1} {\bf \Pi}_K{\bf \hat \Pi}^{\tau_1-1}]_{ji} p_{i}(0)\nonumber\\
&=&  \sum_{ji}\!  \sum_{\tau=0}^{\infty}  e^{-x \tau}\!\sum_{0\leq \tau_1 \leq \cdots \leq \tau} \sum_{K}\  [{\bf \Pi}_O {\bf \hat \Pi}^{\tau-\tau_{O+K-1}-1} (e^{- s} {\bf \Pi}_K)\cdots  {\bf \Pi}_O {\bf \hat \Pi}^{\tau_2-\tau_1-1}  (e^{- s}{\bf \Pi}_K){\bf \hat \Pi}^{\tau_1-1}]_{ji} p_{i}(0)
\label{zsx}
\end{eqnarray}
\end{widetext}
where, apart from those sums made explicit, we are in fact also summing over all possible orders of occurrence of the $K$ and $O$ transitions contained in ${\bf \Pi}_K$ and  $ {\bf \Pi}_O$, respectively. We have split the $e^{- s K}$ factor into $K$ factors of the form  $e^{- s}$, associating each of them to one occurrence of ${\bf \Pi}_K$. 

As all possible values of $K$ are summed over in Eq.~(\ref{zsx}), we can combine the (tilted) transitions $e^{- s} {\bf \Pi}_K$ and those included in ${\bf \hat \Pi}$ --- we thus put together every transition that does not contribute to $O$, including the $s$-tilting in those contributing to $K$. We are in a situation analogous to that of the first line of Eq.~(\ref{zx}), but replacing ${\bf \tilde \Pi}$ by ${\bf \hat \Pi} + e^{-s}  {\bf \Pi}_K$. Then, by splitting the $e^{-x \tau}$ on the different operators, and finally summing over all possible values of the intermediate times $\tau_1, \tau_2, \ldots$, just as we did  in our discussion of the $x$-ensemble, we obtain $Z_O(s,x) = \sum_{ji} [({\bm \Pi}^{sx})^O]_{ji} p_{i}(0)$, with a tilted operator
\begin{equation}
{\bm \Pi^{sx}} = {\bf \Pi}_O (e^{x} -{\bf \hat \Pi}- e^{-s}  {\bf \Pi}_K)^{-1}.
\end{equation}
%which is similar to (\ref{transferx}), but again replacing  ${\bf \tilde \Pi}$ by ${\bf \hat \Pi} + e^{-s}  {\bf \Pi}_K$. 
Our remarks on the convergence of the $x$-ensemble operator also apply in this case. And in fact an example of a region in tilting-parameter space where the sum over $\tau$ does not converge is shown in Fig.~\ref{fig2}.

From $\varphi(s,x) = \lim_{\, O\to\infty} \frac{1}{O} \log{Z_O(s,x)}$ one can as usual obtain mean values, fluctuations and higher order cumulants of $K$ and $\tau$ by taking derivatives. In Section III we consider the mean values
\begin{equation}
 \frac{\langle K\rangle }{O} = -\partial_s \varphi(s,x),\ \  \frac{\langle \tau\rangle }{O} = -\partial_x \varphi(s,x),
\end{equation}
in the limit of large $O$ (which also corresponds to large $K$ and large $\tau$, as both are $O$-extensive). The second derivatives with respect to the same tilting parameter yield the fluctuations, and the cross derivatives are again the correlations between $K$ and $\tau$ for fixed and large $O$.

\section{Doob transform}

The Doob transform allows us to obtain a stochastic matrix giving rise to the same statistics as the tilted operator (which is not stochastic) for some observables of interest ($O$, $\tau$, etc.). From the resulting transition matrix ${\bf \Pi_\textrm{Doob}}$ we can obtain the transition probabilities (link weights of the resulting weighted networks) that give rise to such statistical behavior in the long-time limit. 

In Figs.~\ref{fig1}, \ref{fig2} and \ref{fig3}  we show the Doob-transformed dynamics corresponding to some observable statistics of interest (as given by some values of the tilting parameters). More precisely, we display probability fluxes, which are the product of the Doob-transform transition probabilities and the stationary state distribution ${\bf \Pi_\textrm{Doob}} {\bm p}^{\text st}$, where ${\bm p}^{\text st}$ is the vector whose components are the stationary probabilities associated with the different nodes. While the entries of  ${\bf \Pi_\textrm{Doob}}$ are transition probabilities, $(\Pi_\textrm{Doob})_{ji} p_i^{\text st}$ is the joint probability of occupying node $i$ in a given time step and node $j$ in the next one, which is the probability flux of $j\leftarrow i$. The sizes of the colored arrows in panels (c) and (d) of Figs.~\ref{fig2} and \ref{fig3}, as well as the arrows shown for $x\neq 0$ in Fig.~\ref{fig1} (b) and (d), are proportional to these  probability fluxes. 

We next discuss the Doob transform in each of the ensembles reviewed in Appendix B. In the $s$-ensemble, the tilted operator ${\bf \Pi}^s$ is linked to the biased probability $P^s_{\tau}(O)$ through the largest eigenvalue $e^{\theta(s)}$. The left and right eigenvectors associated with it are ${\bf l}_s$ and ${\bf r}_s$ respectively: ${\bf \Pi}^s{\bf r}_s = e^{\theta(s)} {\bf r}_s$,  ${\bf l}_s^T {\bf \Pi}^s= e^{\theta(s)} {\bf l}_s^T$, where $T$ indicates transposition. However the tilted operator is not a proper stochastic matrix ($\sum_j \Pi^s_{ji} \neq 1$), so it does not provide the transition probabilities leading to the fluctuation of interest. In order to convert it into a proper stochastic matrix one resorts to the generalized Doob transform \cite{simon09a,popkov10a,jack2010,chetrite2015}, which is 
%, has a stationary state vector ${\bf ss}$, whose components are the products the corresponding components of ${\bf l_s}$ and  ${\bf r_s}$: $p_i^{\text st} = l_i r_i$ for $i=1,\ldots, N$ ($N$ being the number of nodes). In order to obtain the same observable $O$ statistics as given by the tilted dynamics, but with a proper stochastic operator one resorts to the generalized Doob transform \cite{simon09a,popkov10a,jack2010,chetrite2015}, which is 
\begin{equation}
{\bf \Pi}^s_\textrm{Doob} = e^{-\theta(s)} {\bf \hat{L}}_s {\bf \Pi}^s{\bf \hat{L}}_s^{-1},
\label{doobs}
\end{equation}
where ${\bf \hat{L}}_s$ is a diagonal matrix which has the components of ${\bf l}_s$ along the diagonal. The Doob transform ${\bf \Pi}^s_\textrm{Doob}$ conserves probability: $\sum_j  ({\bf \Pi}^s_\textrm{Doob})_{ji} = e^{-\theta(s)} {\bf l}_s^T {\bf \Pi^s}{\bf \hat{L}}_s^{-1} = {\bf 1}^T $, where ${\bf 1}$ is the unit vector. And it has the stationary distribution ${\bm p}^{\text st}={\bf \hat{L}}_s {\bf r}_s$: ${\bf \Pi}^s_\textrm{Doob}{\bm p}^{\text st} =  e^{-\theta(s)} {\bf \hat{L}}_s {\bf \Pi}^s{\bf \hat{L}}_s^{-1} {\bm p}^{\text st}=  e^{-\theta(s)} {\bf \hat{L}}_s {\bf \Pi}^s{\bf r}_s = {\bf \hat{L}}_s {\bf r}_s = {\bm p}^{\text st}$.

The Doob transform can be generalized to the $ss$-ensemble in a straighforward manner: in Eq.~(\ref{doobs}) the matrices and the SCGF are now functions of both $s_1$ and $s_2$, but otherwise the expression remains intact,
\begin{equation}
{\bf \Pi}_\textrm{Doob}^{s_1s_2} = e^{-\theta(s_1,s_2)} {\bf \hat{L}}_{s_1 s_2} {\bf \Pi}^{s_1 s_2}{\bf \hat{L}}_{s_1 s_2}^{-1},
\label{doobss}
\end{equation}
 and satisfies analogous properties of stochasticity and having the same stationary distribution as the tilted dynamics.

The $x$-ensemble tilted operator ${\bm \Pi^x}$ satisfies  ${\bm \Pi^x}{\bf r}_x = e^{\varphi(x)} {\bf r}_x$,  ${\bf l}_x^\textrm{T} {\bm \Pi}^x= e^{\varphi(x)} {\bf l}_x^\textrm{T}$, where ${\bf l}_x$ and ${\bf r}_x$ are again the eigenvectors associated with the maximum eigenvalue $e^{\varphi(x)}$.  In this case, the Doob transform is
\begin{equation}
{\bf \Pi}_\textrm{Doob}^x = e^{-x} {\bf \hat{L}}_x {\bf \Pi}^{s=\varphi(x)}{\bf \hat{L}}_x^{-1},
\label{doobx}
\end{equation}
where ${\bf \hat{L}}_x$ is a diagonal matrix which has the components of ${\bf l}_x$ along the diagonal. In fact this is just a rewriting of Eq.~\ref{doobs} in terms of $x$-ensemble parameters, based on the ensemble equivalence whereby $\theta(s) = x$ and $\varphi(x) = s$ when the left eigenvectors in both ensembles are equal  \cite{budini2014,garrahan2017}.

The Doob transform of the $sx$-ensemble is written in terms of the left eigenvector of ${\bm \Pi^{sx}}$ associated with the eigenvalue $e^{\varphi(s,x)}$, ${\bf l}_{sx}$. More precisely, on its rearrangement as a diagonal matrix  ${\bf \hat{L}}_{s x}$. It is given by
\begin{equation}
{\bf \Pi}^{s x}_\textrm{Doob} = e^{-x}  {\bf \hat{L}}_{s x}{\bf \Pi}^{s, s^\prime=\varphi(s,x)} {\bf \hat{L}}_{s x}^{-1}.
\label{doobsx}
\end{equation}
This is just a rewriting of Eq.~\ref{doobss} in terms of $sx$-ensemble parameters, based on the ensemble equivalence whereby $\theta(s,s^\prime) = x$ and $\varphi(s,x) = s^\prime$ when the left eigenvectors in both ensembles are equal. These relations are like those between the $s$- and the $x$-ensembles fixing one of the $s$-variables (namely, $s$, which plays the role of $s_1$).

It seems pertinent to conclude by highlighting a connection between the SRW and the MERW that has been recently unveiled by means of the Doob transform. It turns out that the MERW is the Doob transform of the $s$-biased SRW for the observable $\sum_{i} \log k^\text{out}_i$, where the sum is taken over the nodes visited in a walk, with tilting parameter $s=-1$ \cite{coghi2019}. While in the highly symmetric case of regular networks both random walks are equivalent, important qualitative differences in the trajectories are observed already in the presence of small deviations from regularity, including localization effects \cite{burda2009}. In a large-deviation framework, these are a consequence of biasing the SRW in such a way that the walker favors visiting nodes with a larger (logarithm of the) degree. Dynamical phase transitions towards localized states in biased SRWs on networks are more generally explored in Ref.~\cite{debacco2016}.

\section{Large deviations of a random walk on a directed ring with a shortcut}

In the case of the random walk on a directed ring with a shortcut studied in the first part of Section III.A, one can analytically obtain the SCGF of $x$-ensemble $\varphi(x)$ from the large deviations of the probability distribution of the sample mean of the cyclic-walk length $\ell$. 

A cyclic walk that ends at the starting node is performed $M$ times in succession, with cycle lengths $\ell^{(1)}, \ell^{(2)},\ldots \ell^{(M)}$, where each $\ell^{(i)}$ is a Bernoulli trial which takes the values $\ell_p$ and $\ell_q$ with probabilities $p$ and $q=1-p$, respectively. Let $n_p$ be a random variable that gives the fraction of cycles of length $\ell_p$ that occur in the sequence, whose probability distribution takes the following binomial distribution 
\begin{equation}
P_M(n_p) = \frac{M!}{(n_p M)! [(1-n_p)M]!}\, p^{n_p M} (1-p)^{(1-n_p)M}
\end{equation}
For large $M$ a straightforward application of Stirling's approximation, $M! \sim M^M e^{-M}$, yields the large-deviation form $P_M(n_p) \sim e^{-M I(n_p)}$, where $I(n_p) = D_\textrm{KL}(n_p|| p) = n_p \log[n_p/p] + (1-n_p) \log[(1-n_p)/(1-p)]$ is the Kullback-Leibler divergence between a Bernoulli trial with probability $p$ and one with probability $n_p$. Such rate function $I(n_p)$ achieves its minimum (zero) for $n_p = p$, which corresponds to the mean value, $\langle n_p \rangle=p$. Notice that for small fluctuations around the mean, i.e. $|n_p-p|\sim {\cal O}(1/\sqrt{M})$, $I(n_p)$ can be approximated up to second order as $I(n_p)\approx (n_p-p)^2/(2p(1-p))+{\cal O}[(n_p-p)^3]$, which corresponds to the central limit theorem prediction.

As we are interested in the statistics of $\ell = M^{-1} \sum_{i=1}^M \ell^{(i)}$, which can be rewritten as $\ell = n_p \ell_p + (1-n_p) \ell_q$, we simply change variables. The probability distribution of $\ell$ thus acquires the large-deviation form
\begin{equation}
P_M(\ell)  \sim e^{-M I(\ell)}.
\end{equation} 
Here, $I(\ell)$ is just $I(n_p =  (\ell - \ell_q)/(\ell_p - \ell_q))$, and the Jacobian factor of the change of variable has been neglected as it does not alter the dominant exponential form of the distribution for large $M$.

 The number of times $M$ that the random walker returns to the starting node is fixed, but the duration of the trajectory is fluctuating. Therefore, we study the random-walk trajectories in the $x$-ensemble. The partition function of the $x$-ensemble distribution $P_M^x(\ell) = Z_M^{-1}(x) e^{-x M \ell} P_M(\ell)$ takes the form 
\begin{equation}
Z_M(x) =\sum_\ell  e^{-x M \ell} P_M(\ell) \sim e^{M \varphi(x)},
\end{equation}
where the SCGF $\varphi(x) = -\textrm{min}_\ell [x \ell + I(\ell)]$, which is the Legendre transform of the rate function $I(\ell)$, is obtained from a saddle-point approximation. 

We then minimize $x \ell + I(\ell)$, which amounts to finding the value $\ell^*$ such that $I^\prime(\ell^*) = -x$, so we can write  $\varphi(x) = -x \ell^*- I(\ell^*)$. The minimum is found for
\begin{equation} 
\ell^* = \frac{p e^{-x (\ell_p - \ell_q)} \ell_p + (1-p) \ell_q}{p e^{-x (\ell_p -\ell_q)}  + (1-p) }.
\end{equation} 
In terms of $p_x =  p e^{-x (\ell_p - \ell_q)}/[p e^{-x (\ell_p - \ell_q)}  + (1-p)]$, $\ell^* = p_x \ell_p + (1- p_x) \ell_q$, which is the mean length for a process that avoids taking the shortcut with probability $p_x$. The SCGF is then
\begin{equation} 
\varphi(x)=-x [p_x \ell_p+ (1-p_x) \ell_q]- D_\textrm{KL}(p_x|| p),
\end{equation}
where $D_\textrm{KL}(p_x|| p)$ is the Kullback-Leibler divergence between a Bernoulli trial with probability $p$ and one with probability $p_x$. This expression can be simplified so as to yield the simple form that appears in the main text,
\begin{equation} 
\varphi(x)= -x \ell_q + \log[p\, e^{-x(\ell_p -\ell_q)} + (1-p) ].
\end{equation}
As expected, the same result is obtained from the analytical calculation of the largest eigenvalue of the $x$-ensemble transfer operator for the directed ring (\ref{transferx}).

The first derivative of the SCGF yields the average cyclic path length
\begin{equation}
 \langle \ell \rangle_x = -\varphi^\prime(x) = \ell_q + (\ell_p - \ell_q)\frac{p}{p+(1 - p) e^{x (\ell_p - \ell_q)}}.
\end{equation}
For large absolute values of $x$, $\langle \ell \rangle_x \approx  \ell_q$ if $x>0$, and $\langle \ell \rangle_x \approx  \ell_p$ if $x<0$. This is in agreement with the asymptotic values observed in Fig.~\ref{fig1} (c), and with the intuitively expected result of a strong bias towards shorter or longer paths, respectively. As $x$ appears multiplying the difference between path lengths in the exponent $x (\ell_p - \ell_q)$, a weaker (stronger) bias will be required to approach the asymptotic values in rings where the shortcut starts closer to the beginning (end) of the path around the ring.

\end{document}